\documentclass[a4paper,fleqn]{cas-dc}

\usepackage[numbers]{natbib}

\begin{document}
\let\WriteBookmarks\relax
\def\floatpagepagefraction{1}
\def\textpagefraction{.001}

\shorttitle{Phase-Constrained Self-Supervised MRI Reconstruction}
\shortauthors{Juan Liu et~al.}

\title[mode=title]{Phase-Constrained Self-Supervised Reconstruction and Denoising for Ultra-High-Resolution MRI}

\author[1]{Juan Liu}
% \ead{juan.liu@utsouthwestern.edu}
\author[1,2]{Zhuoheng Liu}
% \ead{zhuoheng.liu@utsouthwestern.edu}
\author[1,2]{Hongbo Wu}
% \ead{hongbo.wu@utsouthwestern.edu}
\author[1]{Xinyue Han}
% \ead{xinyue.han@utsouthwestern.edu}
\author[1,2,3]{Jiaen Liu}
\cormark[1]
\ead{jiaen.liu@utsouthwestern.edu}
\author[1,2,3]{Nian Wang}
\cormark[1]
\ead{nian.wang@utsouthwestern.edu}

\affiliation[1]{organization={Advanced Imaging Research Center, University of Texas Southwestern Medical Center},
    state={Texas}, country={USA}}
\affiliation[2]{organization={Department of Biomedical Engineering, University of Texas Southwestern Medical Center},
    state={Texas}, country={USA}}
\affiliation[3]{organization={Peter O'Donnell Jr. Brain Institute, University of Texas Southwestern Medical Center},
    state={Texas}, country={USA}}
\cortext[1]{Corresponding authors.}

\begin{abstract}
Ultra-high-field gradient-echo (GRE) MRI enables visualization of fine anatomical structures with strong susceptibility contrast. However, these benefits come with substantial signal-to-noise ratio (SNR) penalties from small voxel sizes and rapid phase variation and wrapping due to increased magnetic susceptibility effects. Jointly addressing noise and phase complexity is particularly challenging for quantitative susceptibility mapping (QSM), which is derived from GRE phase and requires noise suppression without blurring fine anatomy or distorting phase information. We developed a measurement-domain self-supervised learning (SSL) framework that jointly reconstructs and denoises complex-valued images from a single noisy acquisition. By partitioning k-space samples into input and held-out subsets, the framework enables training without clean or fully sampled reference data. Another central innovation is an explicit phase-modulation operator incorporated into the unrolled network’s encoding model. Derived from a low-resolution phase estimate, this operator reduces the phase complexity presented to the learned regularizer while maintaining consistency with the acquired complex k-space measurements. We evaluated the framework using ex vivo mouse brain MRI at 15, 18, and 72 $\mu$m isotropic resolution, in vivo mouse brain MRI at 30 $\mu$m isotropic resolution, and in vivo human brain MRI at $0.3 \times 0.3 \times 0.4$ mm$^3$ resolution. Across controlled noise levels, SSL improved image-quality metrics under most evaluated conditions, preserved fine anatomy, and reduced error propagation into $T_2^*$ maps and QSM. Ablation experiments confirmed that phase regularization suppressed localized artifacts associated with rapid phase variation, while in vivo experiments demonstrated the feasibility of noise reduction with structural preservation. These findings support phase-constrained, measurement-domain self-supervision for high-resolution anatomical and quantitative MRI when clean reference data are unavailable.
\end{abstract}

% \begin{highlights}
% \item Self-supervised MRI reconstruction and denoising use a single noisy acquisition.
% \item Phase regularization reduces localized artifacts in complex-valued GRE images.
% \item Fine anatomical structures are preserved, with improved T$_2^*$ maps and QSM.
% \item Validation spans ex vivo and in vivo mouse brains and high-resolution human MRI.
% \end{highlights}

% \begin{Statement of Significance}
% Ultra-high-resolution MRI reveals fine anatomy, but low signal-to-noise ratio and complex gradient-echo phase limit image quality and quantitative reliability. This work introduces a measurement-domain self-supervised framework that jointly reconstructs and denoises a single noisy acquisition without clean training references. Its key innovation integrates phase modulation into an unrolled reconstruction network, reducing phase-related artifacts while maintaining consistency with acquired measurements. Evaluation across mouse and human MRI demonstrates improved structural preservation and downstream $T_2^*$ mapping and quantitative susceptibility mapping. The framework advances reconstruction methodology for settings where repeated acquisitions are impractical, supporting more reliable high-resolution anatomical and quantitative imaging.
% \end{Statement of Significance}


\maketitle

\section{Introduction}

Magnetic resonance imaging (MRI) is a noninvasive modality with excellent soft-tissue contrast and broad applications in neuroimaging \cite{frisoni2010clinical, mahlknecht2010significance, bakshi2008mri}, musculoskeletal imaging \cite{sneag2023mri, vosshenrich2024modern}, cardiac evaluation \cite{rajiah2023cardiac, morales2024present}, and oncology \cite{lagendijk2008mri, ng2023mri}. Increasing spatial resolution can improve the detection of subtle abnormalities and the delineation of cortical layers, subcortical nuclei, and fine vascular structures \cite{feinberg2023next, boulant2024vivo, liu2026mesoscale, serger2026motion, dong2025romer}. In preclinical imaging, resolutions approaching cellular dimensions can reveal microscopic tissue organization that is obscured at conventional resolutions \cite{han2024high, han2026high}. High resolution is also valuable for quantitative MRI, which provides objective measures of tissue properties for disease characterization and longitudinal assessment. T$_2^*$ mapping quantifies effective transverse relaxation, whereas quantitative susceptibility mapping (QSM) estimates magnetic susceptibility; both are sensitive to tissue microstructure and composition. Finer spatial resolution can increase anatomical specificity and reduce partial volume effects in these quantitative maps. Together with the increasing availability of ultra-high-field systems, these benefits are driving demand for MRI at progressively finer spatial scales.

However, high-resolution MRI is constrained by both long acquisition times and low signal-to-noise ratio (SNR) \cite{plenge2012super}. Sampling a larger k-space extent increases scan duration and sensitivity to motion and system instability, whereas smaller voxels contain fewer contributing spins and therefore produce less signal. Parallel imaging \cite{pruessmann1999sense, griswold2002generalized} and compressed sensing \cite{lustig2007sparse, lustig2008compressed} can reduce acquisition time but may compromise fine structural information or increase reconstruction complexity. Conversely, signal averaging can improve SNR only by further extending the acquisition. Thus, acquisition time and SNR impose coupled constraints: reducing one limitation often worsens the other. Overcoming this tradeoff requires reconstruction methods that suppress noise while preserving fine anatomical detail and the accuracy of downstream quantitative maps.

Deep learning has substantially advanced MRI reconstruction, including convolutional and transformer-based models \cite{he2016identity, dosovitskiy2020image, liu2021swin}, physics-informed unrolled networks \cite{sun2016deep, monga2021algorithm}, and diffusion-based solvers \cite{song2021solving, chung2022score}. However, many of these methods require fully sampled or high-quality training targets. Such references are difficult to obtain in high-resolution MRI, where acquisitions are already lengthy and the resulting images remain intrinsically noisy. Repeated measurements to improve SNR further increase acquisition time and susceptibility to motion and scanner drift, limiting the applicability of conventional supervised reconstruction. Self-supervised learning instead derives supervisory signals from the measured data without external labels \cite{liu2021self, he2022masked}. For example, Noise2Void \cite{krull2019noise2void} and Noise2Self \cite{batson2019noise2self} learn to denoise individual images by withholding information at the prediction location. In MRI, self-supervised learning via data undersampling (SSDU) trains reconstruction networks directly from undersampled k-space measurements \cite{yaman2020self}. However, SSDU was developed primarily to address undersampling and does not explicitly target the severe noise encountered in high-resolution MRI.

Recent methods have applied deep learning to MRI reconstruction under noisy conditions, including AUTOMAP for low-field MRI \cite{koonjoo2021boosting}, Noise2Recon for SNR-robust reconstruction \cite{desai2023noise2recon}, k-space Noisier2Noise \cite{millard2023theoretical, moran2020noisier2noise}, and self-supervised reconstruction of low-field lung MRI \cite{chen2026self}. Although promising, these approaches typically rely on supervised or semi-supervised data, image-domain self-supervision, or synthetic noise injection. Image-domain Noise2Noise-like methods may struggle when reconstructed images contain structured artifacts in addition to noise. Moreover, Noisier2Noise further degrades the measurements through synthetic noise injection, which is undesirable when the acquired data already have very low SNR. Other studies have demonstrated that image-domain denoising can improve anatomical visualization and quantitative accuracy \cite{manjon2008mri, veraart2016denoising, zormpas2021accelerating, kang2024self}. For example, Patch2Self and Patch2Self2 improve diffusion MRI denoising, fiber-bundle coherence, and tractography accuracy \cite{fadnavis2020patch2self, fadnavis2024patch2self2}, while Zhang et al. and Quah et al. applied deep learning denoising to improve T$_2^*$ mapping and QSM \cite{zhang2022blip, quah2026complex}. However, because image-domain methods operate after reconstruction, they cannot prevent reconstruction-stage noise amplification or artifacts from propagating into downstream quantitative maps.

Complex-valued gradient-echo (GRE) reconstruction introduces an additional challenge. Susceptibility differences and $B_0$ inhomogeneity can produce strong spatial phase variations, particularly near tissue boundaries and air--tissue interfaces. Although this phase information is essential for applications such as QSM, its complexity can make it difficult for a learned image prior to distinguish anatomical structure from rapid phase changes, resulting in unstable training and localized reconstruction artifacts. An effective reconstruction framework must therefore preserve consistency with the measured complex data while reducing the phase complexity presented to the learned regularizer.

In this work, we propose a measurement-domain self-supervised deep learning framework for joint reconstruction and denoising of ultra-high-resolution MRI directly from noisy, undersampled k-space. The framework simultaneously reconstructs images and suppresses noise without requiring clean reference data or repeated acquisitions. We compare the proposed method with conventional model-based and learning-based approaches under simulated and experimentally acquired low-SNR conditions and evaluate its ability to preserve fine anatomy and improve downstream quantitative imaging.

The main contributions are as follows:
\begin{itemize}
    \item A unified self-supervised framework that jointly reconstructs and denoises MR images from a single noisy acquisition without clean reference data.
    \item A phase-constrained unrolled reconstruction network that reduces artifacts caused by rapid phase variation in complex-valued GRE data.
    \item Extensive validation across spatial resolutions, noise levels, field strengths, and anatomical settings, including evaluation of downstream T$_2^*$ mapping and QSM.
\end{itemize}

\section{Methods}

\subsection{Problem Formulation}

MRI reconstruction aims to recover a spatially resolved image from measurements acquired in k-space and can be formulated as an inverse problem. Let $\mathbf{x}\in\mathbb{C}^{N}$ denote the unknown complex-valued MR image, and let $\mathbf{y}_{\Omega}$ denote the acquired multicoil k-space measurements indexed by the sampling set $\Omega$. The noisy measurement model is

\begin{equation}
    \mathbf{y}_{\Omega} = \mathbf{s}_{\Omega} + \mathbf{n}_{\Omega},
    \qquad
    \mathbf{s}_{\Omega} = \mathbf{E}_{\Omega}\mathbf{x},
    \label{eq:clean_noisy_kspace}
\end{equation}

\noindent where $\mathbf{s}_{\Omega}$ denotes the unknown noise-free measurements, $\mathbf{n}_{\Omega}$ denotes measurement noise, and $\mathbf{E}_{\Omega}=\mathbf{U}_{\Omega}\mathbf{F}\mathbf{C}$ is the encoding operator. Here, $\mathbf{C}$ applies coil sensitivity maps, $\mathbf{F}$ denotes the coilwise Fourier transform, and $\mathbf{U}_{\Omega}$ selects the acquired k-space locations. Noise in raw MRI k-space is typically modeled as complex-valued, additive, zero-mean Gaussian noise. After coil-noise prewhitening, the noise is approximately independent across receiver channels and k-space locations.

Image reconstruction can be formulated as the regularized inverse problem

\begin{equation}
    \hat{\mathbf{x}} =
    \operatorname*{arg\,min}_{\mathbf{x}}
    \left\|
        \mathbf{E}_{\Omega}\mathbf{x}
        -
        \mathbf{y}_{\Omega}
    \right\|_2^2
    +
    \lambda\mathcal{R}(\mathbf{x}),
    \label{eq:regularized_reconstruction}
\end{equation}

\noindent where the first term enforces data consistency between the reconstructed image and the acquired measurements, $\mathcal{R}(\mathbf{x})$ incorporates prior information on the expected image structure, and $\lambda$ controls the strength of regularization. The prior may be explicitly defined, such as sparsity in a transform domain, or implicitly learned by a neural network.

\subsection{Joint Self-Supervised Reconstruction and Denoising Framework}

We extend SSDU \cite{yaman2020self} to jointly reconstruct and denoise images from noisy, undersampled k-space measurements. During training, the acquired sampling set is partitioned into disjoint input and held-out subsets, $\Omega=\Theta\cup\Lambda$. The network reconstructs an image using only measurements indexed by $\Theta$:

\begin{equation}
\hat{\mathbf{x}} = f_{\theta}\left( \mathbf{y}_{\Theta}; \mathbf{E}_{\Theta} \right).
\label{eq:partition_reconstruction}
\end{equation}

The corresponding prediction at the held-out k-space locations is

\begin{equation}
    \hat{\mathbf{s}}_{\Lambda,\theta}
    = \mathbf{E}_{\Lambda} f_{\theta} \left( \mathbf{y}_{\Theta}; \mathbf{E}_{\Theta} \right).
    \label{eq:heldout_prediction}
\end{equation}

Because $\mathbf{y}_{\Lambda}$ is excluded from the network input, the prediction $\hat{\mathbf{s}}_{\Lambda,\theta}$ does not directly depend on the noise realization $\mathbf{n}_{\Lambda}$. Assuming that the held-out noise is conditionally zero mean,

\begin{equation}
\mathbb{E} \left[\mathbf{n}_{\Lambda}\mid\mathbf{s}_{\Omega},\mathbf{y}_{\Theta}\right]
= \mathbf{0},
\label{eq:heldout_noise_assumption}
\end{equation}

\noindent the expected squared-error loss satisfies

\begin{equation}
\begin{aligned}
    &\mathbb{E}
    \left[
        \left\|
            \hat{\mathbf{s}}_{\Lambda,\theta}
            -
            \mathbf{y}_{\Lambda}
        \right\|_2^2
    \right]
    \\
    &={}
    \mathbb{E}
    \left[
        \left\|
            \hat{\mathbf{s}}_{\Lambda,\theta}
            -
            \mathbf{s}_{\Lambda}
        \right\|_2^2
    \right]
    +
    \mathbb{E}
    \left[
        \left\|
            \mathbf{n}_{\Lambda}
        \right\|_2^2
    \right].
\end{aligned}
\label{eq:kspace_self_supervised_risk}
\end{equation}

The second term in Eq.~\eqref{eq:kspace_self_supervised_risk} is independent of the network parameters. Thus,

\begin{equation}
\begin{aligned}
    &\operatorname*{arg\,min}_{\theta}
    \mathbb{E}
    \left[
        \left\|
            \hat{\mathbf{s}}_{\Lambda,\theta}
            -
            \mathbf{y}_{\Lambda}
        \right\|_2^2
    \right]
    \\
    &={}
    \operatorname*{arg\,min}_{\theta}
    \mathbb{E}
    \left[
        \left\|
            \hat{\mathbf{s}}_{\Lambda,\theta}
            -
            \mathbf{s}_{\Lambda}
        \right\|_2^2
    \right].
\end{aligned}
\label{eq:kspace_same_minimizer}
\end{equation}

Equation~\eqref{eq:kspace_same_minimizer} shows that, under the stated assumptions, minimizing the expected squared error against the noisy held-out k-space measurements yields the same population minimizer as training against the unknown noise-free measurements. This result follows from the principle of $\mathcal{J}$-invariance underlying Noise2Self \cite{batson2019noise2self}. The subsets $\Theta$ and $\Lambda$ contain complementary measurements associated with the same underlying image. In short, the k-space partition strategy serves two purposes: it eliminates the need for fully sampled reference data and creates a self-supervised denoising task by requiring the network to predict held-out measurements from complementary noisy samples. 

The acquired k-space set $\Omega$ is randomly repartitioned during training while maintaining the prescribed partition ratio and sampling strategy. Consequently, different measurements are held out across training iterations. This stochastic partitioning prevents the network from becoming specialized to a fixed loss mask and allows the training objective to cover the acquired k-space locations over time. It also encourages the network to recover reproducible image structure rather than fit individual noise realizations. 

Following SSDU, the practical training objective uses a normalized combined $\ell_1$--$\ell_2$ loss. For a reference vector $\mathbf{u}$ and prediction $\mathbf{v}$, the loss is defined as

\begin{equation}
    \mathcal{L}_{\ell_1\text{-}\ell_2}
    (\mathbf{u},\mathbf{v})
    =
    \frac{
        \left\|
            \mathbf{u}-\mathbf{v}
        \right\|_2
    }{
        \left\|\mathbf{u}\right\|_2
    }
    +
    \frac{
        \left\|
            \mathbf{u}-\mathbf{v}
        \right\|_1
    }{
        \left\|\mathbf{u}\right\|_1
    }.
    \label{eq:normalized_l1_l2}
\end{equation}

Normalization reduces sensitivity to variations in measurement scale and in the number of held-out samples. The exact unbiased-risk equivalence in Eq.~\eqref{eq:kspace_self_supervised_risk} applies to the unnormalized squared $L_2$ loss. The normalized combined $\ell_1$--$\ell_2$ loss does not, in general, satisfy the same exact quadratic risk decomposition. The squared-error analysis therefore provides the theoretical motivation for measurement-domain self-supervision, while the normalized combined loss is used as a practical and robust optimization objective.

\begin{figure*}
\centering
\includegraphics[width=.9\textwidth]{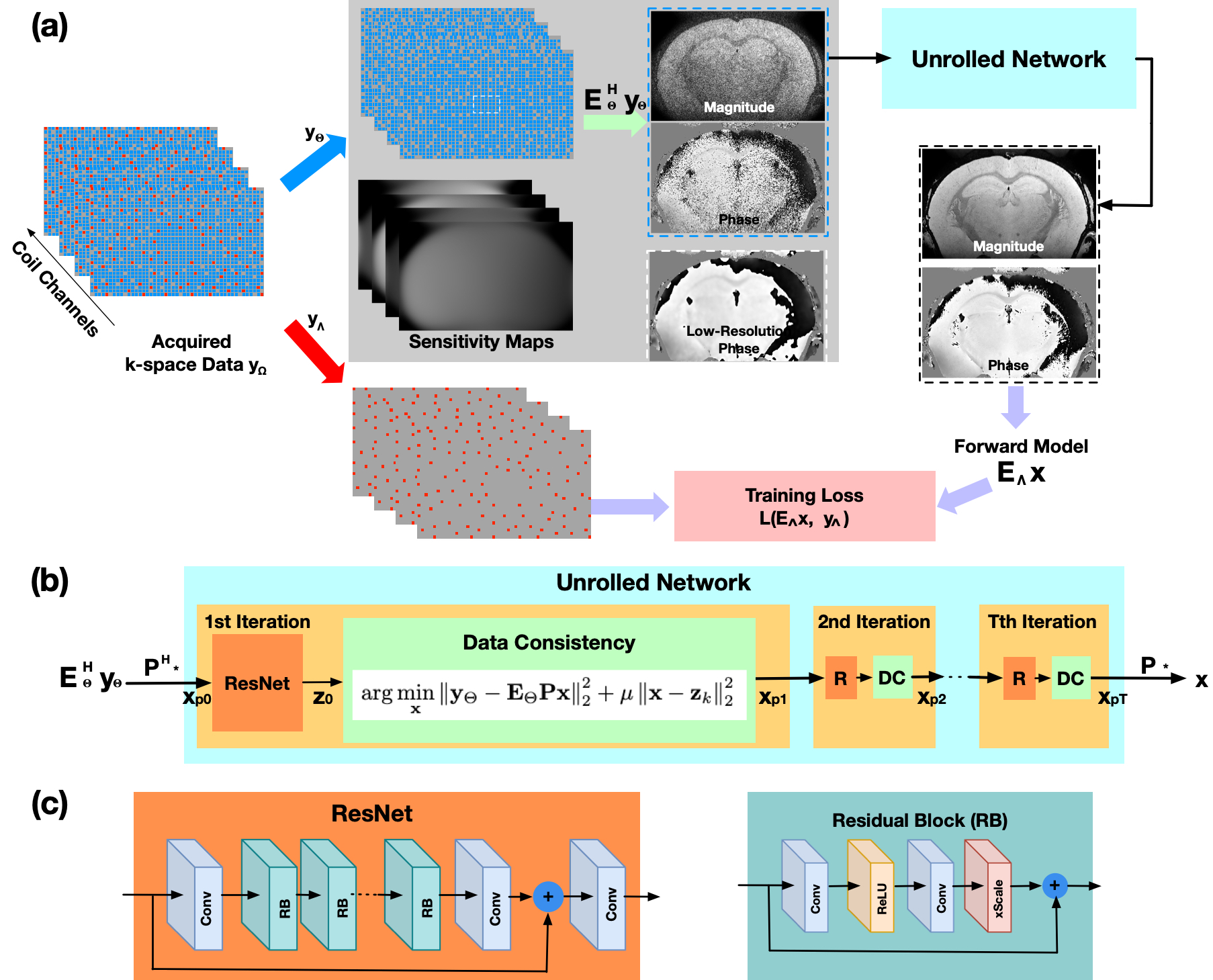}
\caption{Schematic of the proposed self-supervised reconstruction and denoising framework. During training, the acquired k-space set $\Omega$ is partitioned into input samples $\Theta$ and held-out samples $\Lambda$. The zero-filled input $\mathbf{y}_{\Theta}$ is mapped to image space by $\mathbf{E}_{\Theta}^{H}$, while its central k-space region (white dashed box) is used to estimate the low-resolution phase $\mathbf{P}$. The unrolled network alternates learned regularization and data-consistency updates. Its output is projected onto $\Lambda$ using $\mathbf{E}_{\Lambda}$ and compared with $\mathbf{y}_{\Lambda}$ to compute the self-supervised training loss. During inference, all acquired samples in $\Omega$ are used. The network processes the real and imaginary components of the complex-valued images; magnitude and phase images are shown only for visualization.}
\label{fig:ssl_workflow}
\end{figure*}

\subsubsection{SSL Unrolled Reconstruction Network}

The SSL reconstruction network unrolls a variable-splitting algorithm that alternates learned regularization and data consistency. For available k-space samples $\mathcal{S}\subseteq\Omega$, the objective is

\begin{equation}
(\hat{\mathbf{x}},\hat{\mathbf{z}})=
\operatorname*{arg\,min}_{\mathbf{x}, \mathbf{z}}
\left\| \mathbf{y}_{\mathcal{S}} - \mathbf{E}_{\mathcal{S}}\mathbf{x} \right\|_2^2
+ \mu \left\| \mathbf{x} - \mathbf{z} \right\|_2^2
+ \mathcal{R}(\mathbf{z}),
\end{equation}

\noindent where $\mathbf{x}$ is the reconstructed image, $\mathbf{z}$ is an auxiliary variable, $\mathcal{R}$ is the image prior, and $\mu>0$ couples the two variables. The regularization update is represented by a neural network,

\begin{equation}
\mathbf{z}^{(i)} = D_{\theta}
\left(\mathbf{x}^{(i-1)}\right).
\label{eq:learned_regularization_update}
\end{equation}

\noindent The quadratic data-consistency update is

\begin{equation}
\mathbf{x}^{(i)}=
\operatorname*{arg\,min}_{\mathbf{x}}
\left\|\mathbf{y}_{\mathcal{S}}-\mathbf{E}_{\mathcal{S}}\mathbf{x}\right\|_2^2
+\mu\left\|\mathbf{x}-\mathbf{z}^{(i)}\right\|_2^2,
\label{eq:data_consistency_update}
\end{equation}

\noindent and is solved by conjugate gradients. The coupling parameter $\mu$ is initialized from the estimated noise level and optimized during training. The network is initialized by the adjoint reconstruction

\begin{equation}
\mathbf{x}^{(0)} = \mathbf{E}_{\mathcal{S}}^{H} \mathbf{y}_{\mathcal{S}}.
\label{eq:initial_image}
\end{equation}

These updates are repeated for a fixed number of iterations. During training, $\mathcal{S}=\Theta$ and the output is projected onto $\Lambda$ for loss evaluation; during inference, $\mathcal{S}=\Omega$ so that all acquired measurements are used.

\subsubsection{Phase Regularization} 

Gradient-echo images can exhibit rapid spatial phase variations near susceptibility boundaries, including air--tissue interfaces. In preliminary experiments, these variations produced localized reconstruction artifacts. To mitigate this problem, a smooth phase estimate was derived from central k-space and incorporated into the encoding model. We formulate the optimization as

\begin{equation}
    \hat{\mathbf{x}} =
    \operatorname*{arg\,min}_{\mathbf{x}}
    \left\|
        \mathbf{U}_{\Omega}\mathbf{F}\mathbf{C}\mathbf{P}\mathbf{x}
        -
        \mathbf{y}_{\Omega}
    \right\|_2^2
    +
    \lambda\mathcal{R}(\mathbf{x}),
    \label{eq:phase_regularized_reconstruction}
\end{equation}

\noindent where $\mathbf{P}$ is a diagonal unit-magnitude operator representing the low-resolution phase induced predominantly by $B_0$ inhomogeneity. During training, $\mathbf{P}$ is estimated from the central 1\% of the k-space samples in $\mathbf{y}_{\Theta}$. Incorporating $\mathbf{P}$ into the encoding model preserves phase consistency with the measurements while reducing the residual phase variation represented by $\mathbf{x}$. The denoiser $D_{\theta}$ can therefore focus on image structure and noise rather than modeling the full field-induced phase.

\subsubsection{Phase-Regularization Ablation Design}

The contribution of phase regularization was evaluated by training otherwise identical SSL reconstructions with and without $\mathbf{P}$ using the 15-$\mu$m isotropic single-echo GRE ex vivo mouse brain dataset presented below as Dataset 1. Reconstruction quality was assessed using normalized root-mean-square error (NRMSE), peak signal-to-noise ratio (PSNR), structural similarity index measure (SSIM), and the spatial distribution of residual artifacts.

Fig.~\ref{fig:ssl_workflow} summarizes the complete training and inference workflow, including k-space partitioning, phase estimation, unrolled reconstruction, and held-out loss evaluation.

\subsection{Datasets and Experimental Setup}

Table~\ref{tab:dataset_overview} summarizes the imaging parameters, experimental purposes, and outputs of the five primary datasets used in this study.

\begin{table*}[width=\textwidth,cols=6,pos=t]
\caption{Overview and imaging parameters of the five primary MRI datasets.}
\label{tab:dataset_overview}
\centering
\begin{tabular*}{\tblwidth}{@{\extracolsep{\fill}}
    >{\raggedright\arraybackslash}p{\dimexpr.13\textwidth-2\tabcolsep\relax}
    *{5}{>{\raggedright\arraybackslash}p{\dimexpr.174\textwidth-2\tabcolsep\relax}}@{}}
\toprule
Parameter & \shortstack{Dataset 1\\Ex vivo mouse} & \shortstack{Dataset 2\\Ex vivo mouse} & \shortstack{Dataset 3\\Ex vivo mouse} & \shortstack{Dataset 4\\In vivo mouse} & \shortstack{Dataset 5\\In vivo human} \\
\midrule
Sequence & Single-echo GRE & Dual-echo GRE & Dual-echo GRE & Dual-echo GRE & Dual-echo GRE \\
Field strength & 9.4 T & 9.4 T & 9.4 T & 9.4 T & 7 T \\
Matrix size & $480 \times 756 \times 480$ & $210 \times 160 \times 100$ & $850 \times 640 \times 400$ & $240 \times 400 \times 160$ & $800 \times 600 \times 80$ \\
FOV (mm$^3$) & $7.2 \times 11.34 \times 7.2$ & $15.12 \times 11.52 \times 7.2$ & $15.3 \times 11.52 \times 7.2$ & $7.2 \times 12.0 \times 4.8$ & $240 \times 180 \times 32$ \\
Spatial resolution & 15 $\mu$m isotropic & 72 $\mu$m isotropic & 18 $\mu$m isotropic & 30 $\mu$m isotropic & $0.3 \times 0.3 \times 0.4$ mm$^3$ \\
TE (ms) & 15.7 & 8.0, 19.9 & 8.0, 19.9 & 10.0, 16.5 & 18.0, 39.0 \\
TR (ms) & 85 & 110 & 105 & 100 & 74 \\
Flip angle & $30^\circ$ & $50^\circ$ & $50^\circ$ & $30^\circ$ & $14^\circ$ \\
Receiver bandwidth & 75 kHz & 100 kHz & 100 kHz & 100 kHz & 54 Hz/pixel (43.2 kHz) \\
Repetitions & NEX = 1 & NEX = 1 & NEX = 1 & NEX = 1 & 1 \\
Parallel imaging & None & None & None & None & $2 \times 1$ \\
Acquisition time & $\sim$13 h/repetition &  23 min & 6.0 h & 53 min & 35.5 min \\
\midrule
Experimental purpose & Single-repetition reconstruction and denoising using the averaged three-repetition acquisition as a reference & Robustness and quantitative accuracy under controlled, simulated low-SNR conditions & High-resolution multi-echo anatomical and quantitative imaging & In vivo reconstruction and preservation of cerebral vascular structures & Accelerated human reconstruction and quantitative mapping under original and noise-augmented conditions \\
Image outputs & Magnitude image and QSM & Magnitude images, T$_2^*$ map, and QSM & Magnitude images, T$_2^*$ map, and QSM & Magnitude images, minimum intensity projections, and vessel segmentation & Magnitude images, T$_2^*$ map, and QSM \\
\bottomrule
\end{tabular*}
\end{table*}

\subsubsection{\texorpdfstring{Dataset 1: 15 $\mu$m Single-Echo T$_2^*$-Weighted GRE Imaging of Ex Vivo Mouse Brains}{Dataset 1: 15-micrometer Single-Echo T2*-Weighted GRE Imaging of Ex Vivo Mouse Brain}}

All animal experiments were conducted in accordance with guidelines approved by the local Institutional Animal Care and Use Committee (IACUC). Mice were deeply anesthetized and subsequently euthanized by transcardial perfusion fixation \cite{wang2019probing}. Five ex vivo mouse brains were imaged using a single-echo gradient-echo (GRE) sequence on a 9.4 T Bruker BioSpec system (Bruker BioSpin, Ettlingen, Germany) equipped with a four-channel cryoprobe radiofrequency (RF) coil. The imaging parameters are listed in Table~\ref{tab:dataset_overview}. One brain was imaged with three repetitions, resulting in a proportionally longer total scan time.

Raw k-space was processed using BART \cite{uecker2015berkeley} to estimate coil sensitivities with ESPIRiT \cite{uecker2014espirit} and perform coil-noise prewhitening. Seven approaches were evaluated: conventional inverse FFT with coil combination, Fermi-window k-space apodization \cite{bernstein2001effect}, TV-regularized reconstruction \cite{rudin1992nonlinear}, phase-demodulated nonlocal means (NLM) \cite{buades2011non}, Neighbor2Neighbor \cite{huang2021neighbor2neighbor}, image-domain Noise2Noise using repeated acquisitions, and the proposed method. Noise2Noise was trained using independently reconstructed repetitions. For the proposed method, an inverse FFT along the readout direction produced a hybrid-space representation, and the network was trained on nine-slice 2D slabs to reduce memory use while retaining spatial context.

The averaged three-repetition acquisition served as the reference. Performance was evaluated using NRMSE, PSNR, SSIM, and magnitude error maps. QSM was generated using Laplacian phase unwrapping \cite{schofield2003fast}, V-SHARP background-field removal \cite{ozbay2017comprehensive}, and nonlinear dipole inversion \cite{polak2020nonlinear}.

\subsubsection{\texorpdfstring{Dataset 2: 72 $\mu$m Dual-Echo T$_2^*$-Weighted GRE Imaging of Ex Vivo Mouse Brains}{Dataset 2: 72-micrometer Dual-Echo T2*-Weighted GRE Imaging of Ex Vivo Mouse Brains}}

Nine ex vivo mouse brains were imaged using a dual-echo GRE sequence on a 9.4 T Bruker BioSpec system equipped with a four-channel cryoprobe RF coil. The imaging parameters are listed in Table~\ref{tab:dataset_overview}. This dataset was used to generate controlled low-SNR conditions by adding noise to the acquired k-space data, while the original acquisition served as a high-SNR reference. This design enabled quantitative evaluation across multiple noise levels without requiring an independently acquired, noise-free high-resolution reference, which is difficult to obtain in practice.

Raw k-space was processed using BART with ESPIRiT coil-sensitivity estimation and coil-noise prewhitening. Conventional inverse FFT reconstructions from the original k-space served as quantitative references. Independent complex Gaussian noise was added to generate target SNR values of 25, 20, 15, 10, and 7, compared with a mean original SNR of 72. SNR was calculated from first-echo magnitude images reconstructed by inverse FFT and root-sum-of-squares (RSS) coil combination as the mean brain signal divided by the background-noise standard deviation. The brain signal was measured in the central slice to limit receive-sensitivity variation, and noise was measured outside the sample. The minimum SNR of 7 approximated that of the 15 $\mu$m Dataset 1 acquisitions.

Seven methods were evaluated: conventional inverse FFT with coil combination, Fermi-window k-space apodization, NLM, TV-regularized reconstruction, two learning-based denoisers, and SSL. The learning-based denoisers were Neighbor2Neighbor and Noise2Noise. Noise2Noise was trained on paired image reconstructions generated from two independent k-space noise realizations at each SNR. Both denoisers used $64 \times 64 \times 64$ image patches. For SSL, an inverse FFT along the readout direction produced hybrid-space data, and the network was trained on 2D slabs containing seven adjacent slices.

Following image reconstruction, T$_2^*$ maps and QSM were computed from the dual-echo data. The reconstructed magnitude images, T$_2^*$ maps, and QSM were compared with their corresponding references using NRMSE, PSNR, and SSIM. For the dual-echo magnitude images, metric values were calculated separately for the two echoes and subsequently averaged for statistical analysis.

\subsubsection{\texorpdfstring{Dataset 3: 18 $\mu$m Dual-Echo T$_2^*$-Weighted GRE Imaging of Ex Vivo Mouse Brains}{Dataset 3: 18-micrometer Dual-Echo T2*-Weighted GRE Imaging of Ex Vivo Mouse Brains}}

Fifteen ex vivo mouse brains were imaged using a dual-echo GRE sequence on a 9.4 T Bruker BioSpec system equipped with a four-channel cryoprobe RF coil. The imaging parameters are listed in Table~\ref{tab:dataset_overview}. This dataset was used to evaluate the proposed method for high-resolution anatomical imaging, T$_2^*$ mapping, and QSM.

The raw k-space data were processed using the same preprocessing pipeline described for Dataset 2, including ESPIRiT coil-sensitivity estimation and coil-noise prewhitening. The images were reconstructed using two methods: conventional inverse FFT followed by coil combination and the proposed SSL. Following image reconstruction, T$_2^*$ maps and QSM were generated for downstream evaluation.

\subsubsection{\texorpdfstring{Dataset 4: Dual-Echo T$_2^*$-Weighted GRE Imaging of In Vivo Mouse Brains}{Dataset 4: Dual-Echo T2*-Weighted GRE Imaging of In Vivo Mouse Brains}}

Two mice were imaged in vivo on a 9.4 T Bruker BioSpec system equipped with a four-channel phased-array RF coil. The cohort comprised a healthy B6 control mouse and a 5xFAD mouse, a transgenic model of Alzheimer's disease (AD). Animals were maintained under continuous anesthesia with isoflurane. Respiration (approximately $80\pm15$ breaths/min) and external body temperature were monitored during imaging using an MR-compatible small-animal monitoring and gating system (SA Instruments, Stony Brook, NY). A gadolinium (Gd)-based contrast agent was administered via intraperitoneal injection \cite{wang2019probing}. The imaging parameters are listed in Table~\ref{tab:dataset_overview}.

The raw k-space data were processed using BART for ESPIRiT coil-sensitivity estimation and coil-noise prewhitening. MR images were reconstructed using two methods: conventional inverse FFT followed by coil combination and the proposed SSL framework. Minimum intensity projection (MinIP) images were subsequently generated from the reconstructed magnitude images for visualization and qualitative assessment of the cerebral vasculature. Vessel segmentation was performed on MinIP images using a Hessian-based Frangi vesselness filter \cite{frangi1998multiscale}.
Because this experiment included two animals, it was treated as an in vivo feasibility demonstration.

\subsubsection{\texorpdfstring{Dataset 5: Dual-Echo T$_2^*$-Weighted GRE Imaging of the In Vivo Human Brain}{Dataset 5: Dual-Echo T2*-Weighted GRE Imaging of the In Vivo Human Brain}}

The experiment was conducted on a 7 T MR scanner (MAGNETOM Plus, Siemens Healthineers, Erlangen, Germany) equipped with a 32-channel head-only RF coil (Nova Medical, Wilmington, MA, USA). A healthy participant was recruited under a research protocol approved by the Institutional Review Board. The imaging parameters are listed in Table~\ref{tab:dataset_overview}. MR images were reconstructed using SENSE and SSL without motion correction, followed by T$_2^*$ mapping and QSM using the pipeline described in \cite{van2023effect}. Complex Gaussian noise at twice the estimated acquisition-noise standard deviation was added to k-space, reducing the SNR by approximately 55.3\%. SENSE images and quantitative maps reconstructed from the original data served as references. Noise-augmented reconstructions were evaluated using NRMSE, PSNR, and SSIM. Because this experiment included one participant, it was treated as a feasibility demonstration rather than a cohort-level validation.

\section{Results}

\subsection{Phase Regularization Stabilizes Complex GRE Reconstruction}

\begin{figure}
\centering
\includegraphics[width=\columnwidth]{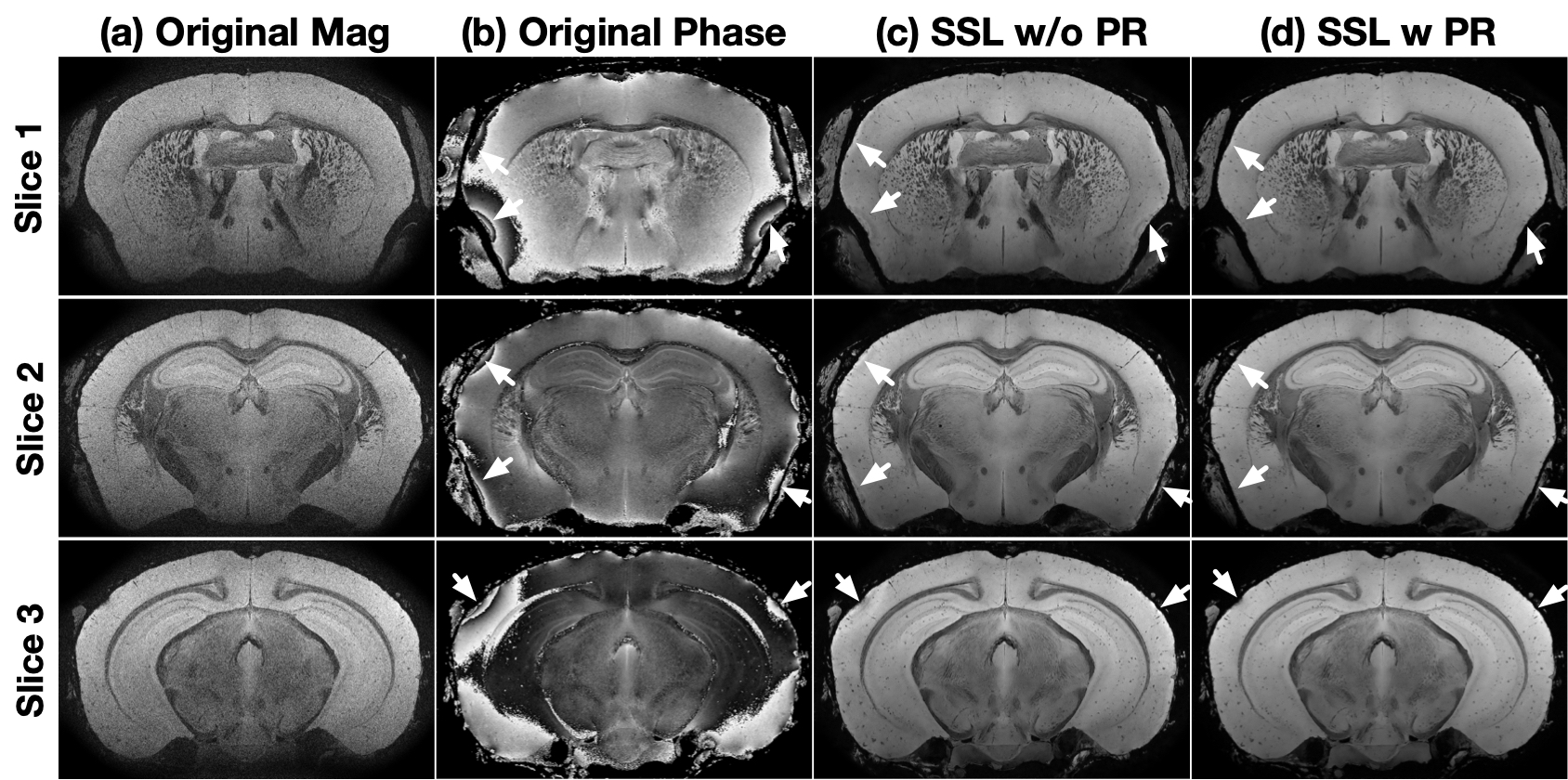}
\caption{Effect of phase regularization on SSL reconstruction of a representative 15-$\mu$m dataset. (a) Original magnitude image, (b) original phase image, (c) SSL without phase regularization, and (d) phase-constrained SSL. White arrows identify localized artifacts associated with rapid phase variation.}
\label{fig:fig_phase_modulation}
\end{figure}

Without phase regularization, rapid phase variations produced localized artifacts in the 15-$\mu$m GRE reconstruction (Fig.~\ref{fig:fig_phase_modulation}). Incorporating the phase estimate reduced NRMSE from 23.40 to 19.61, increased PSNR from 25.45 to 27.84, and increased SSIM from 0.933 to 0.943. The strongest visual improvements occurred in regions of rapid phase variation, supporting phase regularization as an integral component of SSL for complex-valued GRE reconstruction.

\subsection{Single-Acquisition Reconstruction at 15 $\mu$m Resolution}

\begin{figure*}
\centering
\includegraphics[width=.9\textwidth]{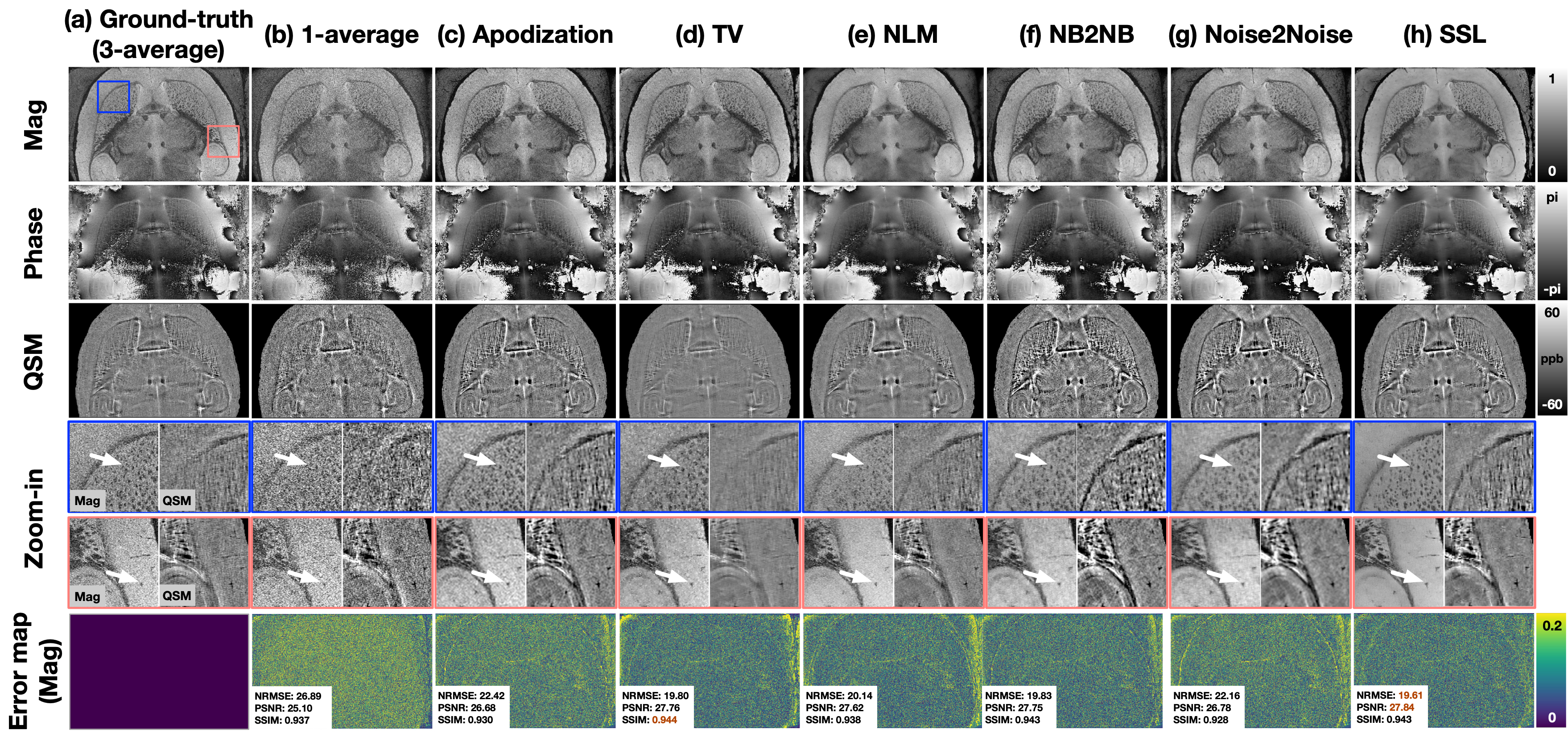}
\caption{Reconstruction and denoising of 15 $\mu$m isotropic ex vivo mouse brain MRI. (a) Three-repetition reference, (b) single-repetition reconstruction, (c) k-space apodization, (d) TV regularization, (e) NLM, (f) Neighbor2Neighbor, (g) Noise2Noise, and (h) proposed SSL. Methods (b)--(h) used the same single-repetition data. SSL best suppressed noise, preserved fine magnitude and QSM features, and achieved the lowest NRMSE and highest PSNR.}
\label{fig:fig_ssl_15um}
\end{figure*}

Fig.~\ref{fig:fig_ssl_15um} compares reconstruction and denoising methods for the 15 $\mu$m isotropic ex vivo mouse brain dataset. The averaged three-repetition reconstruction (a) had higher SNR than the single-repetition reconstruction (b). K-space apodization, TV regularization, and image-domain denoising reduced noise but introduced blurring, residual noise, or loss of fine detail. SSL (h) suppressed noise while preserving magnitude and QSM features. Zoomed views and magnitude error maps relative to the averaged reference showed smaller residual errors with SSL, which achieved the lowest NRMSE (19.61) and highest PSNR (27.84) among the evaluated methods.

\subsection{Robustness to Controlled Measurement Noise}

\begin{figure*}
\centering
\includegraphics[width=.9\textwidth]{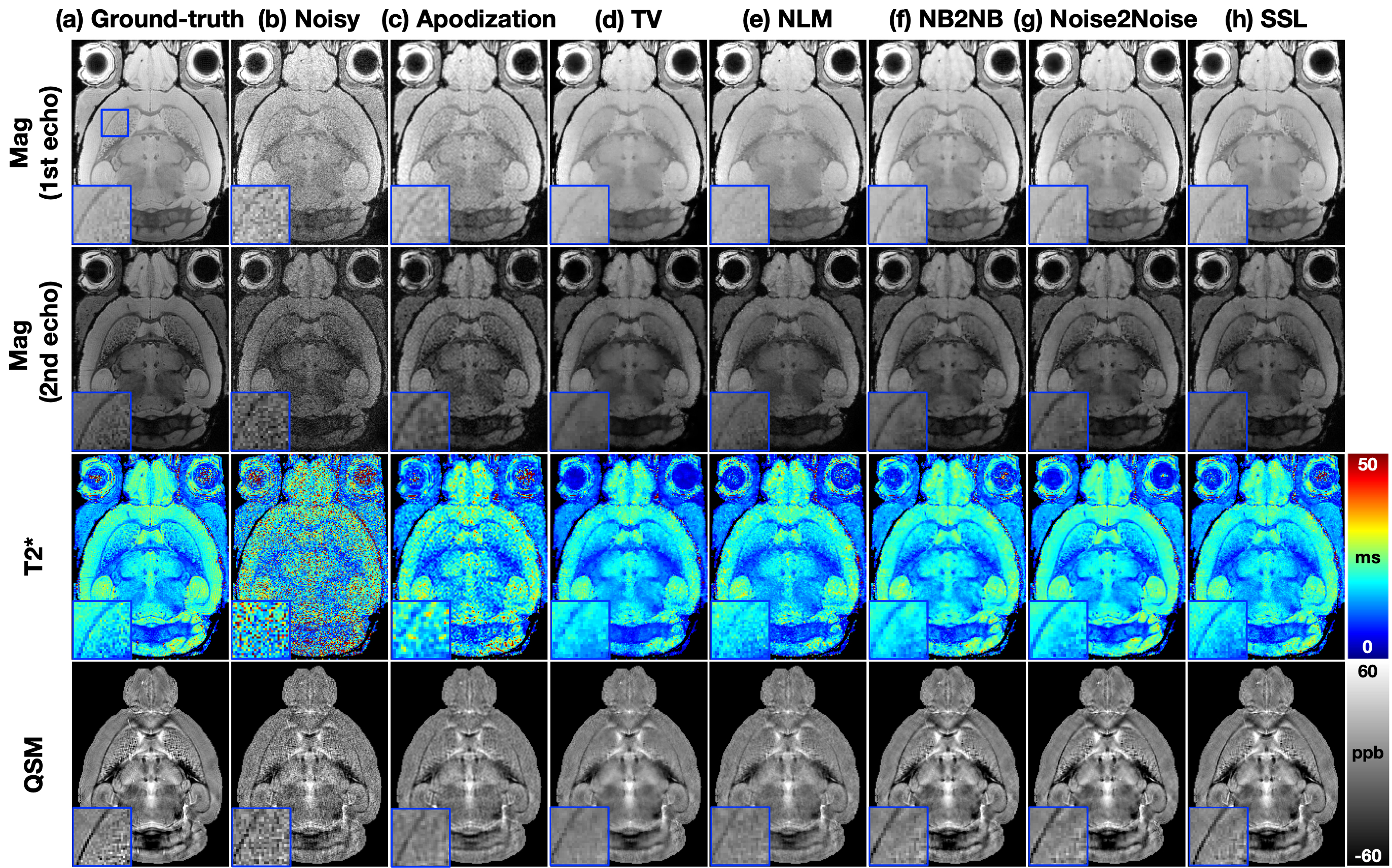}
\caption{Reconstruction and denoising of 72 $\mu$m isotropic ex vivo mouse brain MRI after k-space noise reduced the SNR from 89 to 10. (a) Reference, (b) conventional reconstruction, (c) k-space apodization, (d) TV regularization, (e) NLM, (f) Neighbor2Neighbor, (g) Noise2Noise, and (h) proposed SSL. SSL provided strong noise suppression while preserving anatomical detail and improving the corresponding T$_2^*$ maps and QSM.}
\label{fig:fig_ssl_72um}
\end{figure*}

Fig.~\ref{fig:fig_ssl_72um} compares reconstruction and denoising methods for the 72 $\mu$m isotropic ex vivo mouse brain dataset under simulated low-SNR conditions. Added k-space noise degraded the conventional reconstruction and propagated into the T$_2^*$ maps and QSM. The comparison methods reduced noise but introduced blurring, residual artifacts, or loss of detail. SSL provided stronger noise suppression while better preserving anatomical structures and contrast in both the reconstructed images and quantitative maps.

\begin{figure}
\centering
\includegraphics[width=\columnwidth]{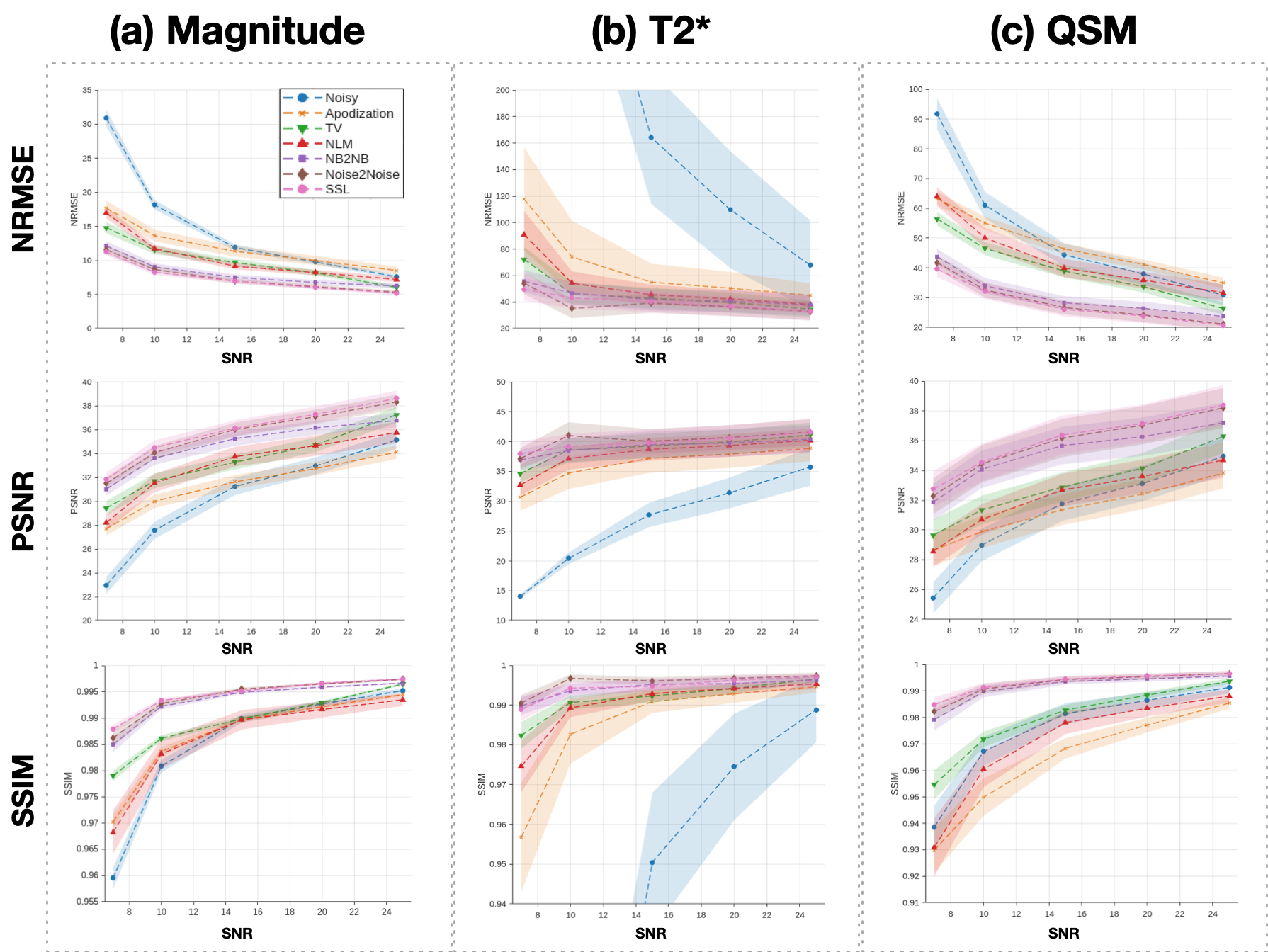}
\caption{NRMSE, PSNR, and SSIM for magnitude images, T$_2^*$ maps, and QSM at simulated SNRs of 7--25, using the original SNR-89 data as the reference. SSL achieved the best metrics across most noise levels and outputs.}
\label{fig:fig_ssl_72um_metrics}
\end{figure}

Fig.~\ref{fig:fig_ssl_72um_metrics} summarizes the quantitative performance of the reconstruction and denoising methods at the five simulated noise levels. NRMSE, PSNR, and SSIM were evaluated for magnitude images, T$_2^*$ maps, and QSM. Across most evaluated SNR conditions and outputs, the proposed SSL method achieved lower NRMSE and higher PSNR and SSIM than the comparison methods. Improvements in both the reconstructed magnitude images and the derived quantitative maps indicate that reducing measurement noise improved downstream quantitative imaging.

\subsection{\texorpdfstring{Dataset 3: 18 $\mu$m Dual-Echo T$_2^*$-Weighted GRE Imaging of Ex Vivo Mouse Brains}{Dataset 3: 18-micrometer Dual-Echo T2*-Weighted GRE Imaging of Ex Vivo Mouse Brains}}

\begin{figure}
\centering
\includegraphics[width=\columnwidth]{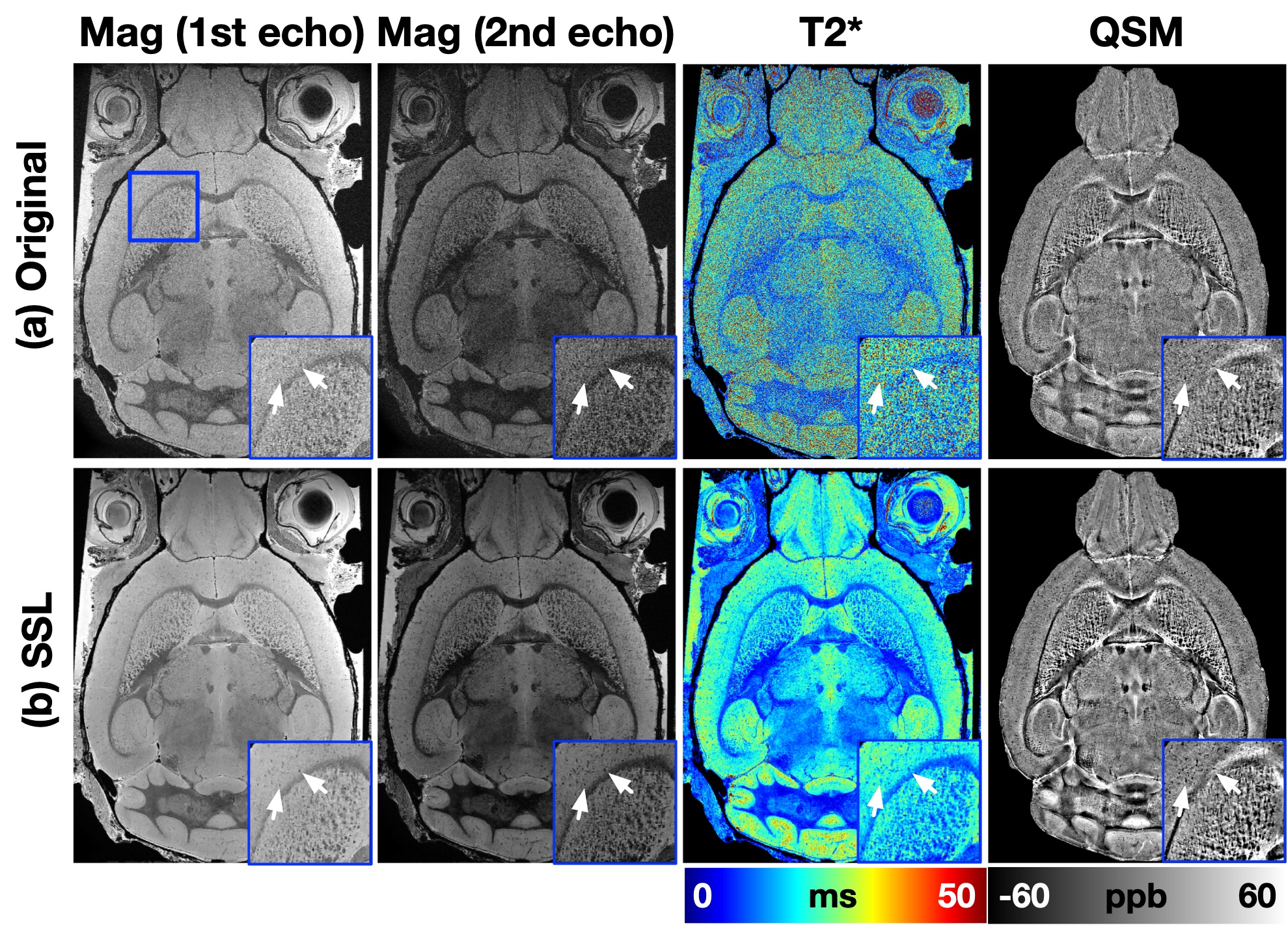}
\caption{Reconstruction and denoising of 18 $\mu$m isotropic ex vivo mouse brain MRI. (a) Conventional reconstruction and (b) proposed SSL. SSL more effectively suppressed noise while preserving fine anatomy and improving the corresponding T$_2^*$ maps and QSM (white arrows).}
\label{fig:fig_ssl_18um}
\end{figure}

Fig.~\ref{fig:fig_ssl_18um} presents representative reconstruction and denoising results from the 18 $\mu$m isotropic dual-echo ex vivo mouse brain dataset. The conventional inverse FFT reconstruction (a) exhibited substantial noise that obscured fine anatomical structures. This noise was also propagated into the corresponding T$_2^*$ and QSM maps, resulting in spatially heterogeneous quantitative estimates. In comparison, the proposed SSL method (b) achieved stronger noise suppression while preserving tissue boundaries and fine structures. The resulting T$_2^*$ and QSM maps exhibited reduced noise propagation and improved preservation of anatomical contrast.

\subsection{\texorpdfstring{Dataset 4: Dual-Echo T$_2^*$-Weighted GRE Imaging of In Vivo Mouse Brains}{Dataset 4: Dual-Echo T2*-Weighted GRE Imaging of In Vivo Mouse Brains}}
\begin{figure}
\centering
\includegraphics[width=\columnwidth]{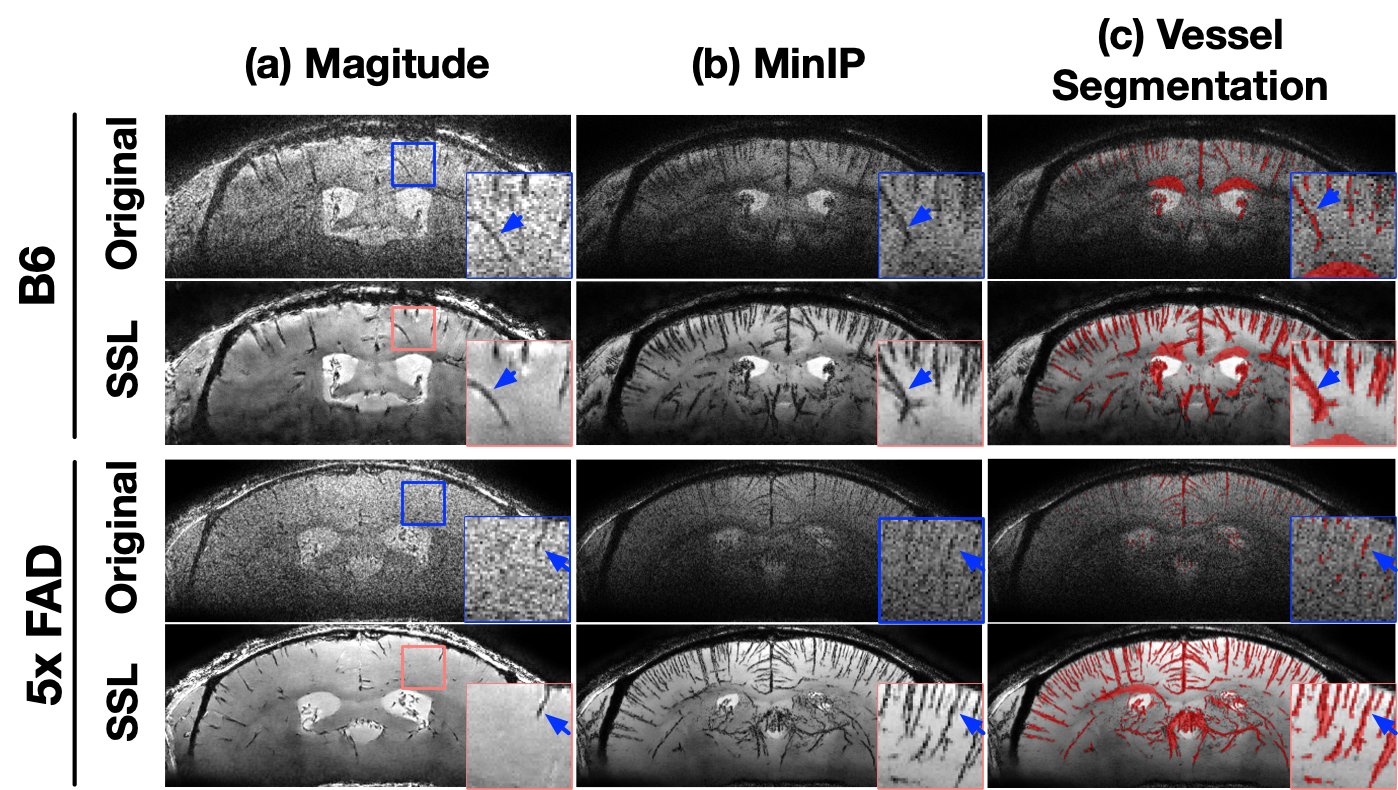}
\caption{In vivo mouse brain MRI at 30 $\mu$m isotropic resolution in healthy B6 and 5xFAD mice. Compared with inverse FFT reconstruction, SSL reduced noise, preserved fine anatomy, improved vascular visibility in first-echo MinIP images, and reduced spurious vessel-segmentation detections (blue arrows).}
\label{fig:fig_ssl_invivo_mouse}
\end{figure}

Fig.~\ref{fig:fig_ssl_invivo_mouse} demonstrates the application of the proposed SSL method to in vivo mouse brain data acquired from one healthy control mouse and one 5xFAD mouse. Compared with conventional inverse FFT reconstruction, the proposed SSL method substantially reduced noise while preserving fine anatomical structures. In the magnitude images, SSL improved tissue contrast and delineation of small structures without introducing marked blurring. MinIP images generated from the SSL reconstructions showed improved visibility and continuity of vascular structures. Vessel segmentation performed on the MinIP images also produced cleaner vessel maps with fewer isolated noise-related detections and more continuous vascular structures than those obtained from the conventional reconstruction. Similar improvements were observed in both the healthy control and 5xFAD datasets.

\subsection{\texorpdfstring{Dataset 5: Dual-Echo T$_2^*$-Weighted GRE Imaging of the In Vivo Human Brain}{Dataset 5: Dual-Echo T2*-Weighted GRE Imaging of the In Vivo Human Brain}}
\begin{figure}
\centering
\includegraphics[width=\columnwidth,height=0.8\textheight,keepaspectratio]{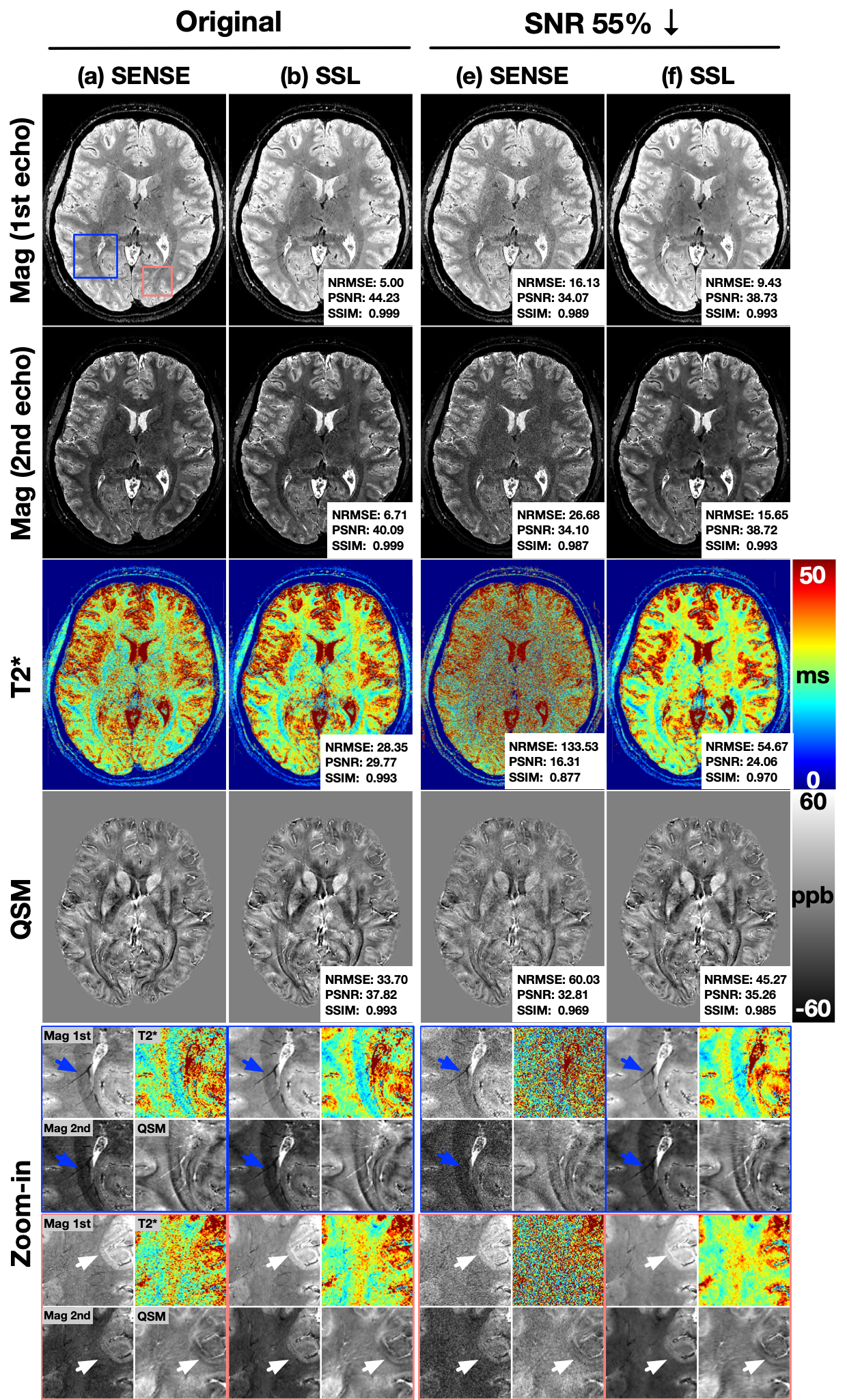}
\caption{In vivo human brain MRI at $0.3 \times 0.3 \times 0.4$ mm$^3$ resolution using original and noise-augmented k-space data. Compared with SENSE, SSL reduced noise, preserved fine anatomy, improved the corresponding T$_2^*$ maps and QSM (blue and white arrows), and achieved better quantitative metrics.}
\label{fig:fig_ssl_invivo_human}
\end{figure}

Fig.~\ref{fig:fig_ssl_invivo_human} demonstrates the application of the proposed SSL method to high-resolution in vivo human brain data acquired from a healthy participant. Under the original acquisition condition, both the SENSE and SSL reconstructions preserved substantial anatomical detail. However, the differences between the two methods became more apparent after additional noise was introduced into the acquired k-space data. Under the reduced-SNR condition, the SENSE reconstruction exhibited greater noise contamination and reduced image quality, obscuring tissue boundaries and fine anatomical features. In contrast, the proposed SSL method effectively suppressed the added noise while preserving structural sharpness and fine anatomical detail.

The improvements observed in the magnitude images extended to the derived T$_2^*$ maps and QSM. Compared with SENSE, SSL reduced noise and visible artifacts while improving anatomical delineation. Using the SENSE reconstruction from the original k-space data as the reference, the addition of noise degraded all evaluated metrics for the magnitude images, T$_2^*$ maps, and QSM. Under the reduced-SNR condition, SSL achieved better quantitative metrics than SENSE, indicating improved preservation of anatomical structure and quantitative information in high-resolution in vivo human MRI.

\section{Discussion}

This study extends SSDU from self-supervised reconstruction to joint reconstruction and denoising in the measurement domain. The central finding is that held-out k-space prediction can provide a denoising signal without clean targets or repeated acquisitions while retaining the physical constraints of the MRI encoding model. Across the evaluated datasets, SSL reduced visible noise and preserved fine structure, with corresponding improvements in T$_2^*$ maps and QSM under controlled low-SNR conditions.

The comparison methods exhibited tradeoffs between noise suppression and spatial resolution. K-space apodization attenuates high-frequency information, while TV regularization and NLM can blur structures through imposed smoothness or self-similarity. The learning-based comparison methods operate after reconstruction, when noise amplification and structured artifacts may already be present. Noise2Noise additionally requires independent noisy pairs. In contrast, SSL uses the measurement model and acquired k-space samples to constrain denoising during reconstruction.

Measurement-domain partitioning provides the statistical basis for joint learning. After prewhitening, noise at disjoint k-space locations is assumed to be approximately zero mean and independent. Excluding held-out samples from the input prevents their noise from directly influencing their predictions. Under the squared-error assumptions in the Methods, noisy and noise-free training risks have the same population minimizer, consistent with $\mathcal{J}$-invariance. Random repartitioning promotes recovery of reproducible structure, and the readout-direction inverse Fourier transform preserves zero-mean noise in the hybrid-space implementation. Data consistency further limits departures from the acquired measurements, although it cannot eliminate all risks of learned bias when the acquired SNR is extremely low.

Phase regularization addresses a GRE-specific failure mode rather than serving as a generic preprocessing operation. By incorporating the low-resolution phase into the forward model, the network remains consistent with the measured complex data while the learned regularizer operates on an image with lower residual phase complexity. The ablation study showed that this formulation reduced localized artifacts and improved all three image-quality metrics. The optimal central k-space kernel for estimating $\mathbf{P}$ remains data dependent, however. Under stronger $B_0$ variation or at longer echo times, an excessively small kernel may inadequately represent the spatial phase variation, whereas an excessively large kernel may absorb anatomical detail. Adaptive phase estimation should therefore be investigated across sequences and field strengths. Phase regularization may be less important for acquisitions with intrinsically smooth phase, such as many spin-echo sequences.

The benefits extended beyond magnitude images. Noise and reconstruction errors propagate through exponential fitting, phase processing, background-field removal, and dipole inversion, potentially amplifying errors in T$_2^*$ maps and QSM. Suppressing noise during reconstruction reduced this propagation in the controlled-noise experiments. Nevertheless, the original-data and averaged reconstructions used here are imperfect references rather than noise-free ground truth, and reconstruction cannot recover information that is overwhelmed by measurement noise. The lowest-SNR results therefore emphasize that adequate acquired signal quality remains essential.

Several limitations remain. The theory assumes additive, complex, zero-mean noise that is independent after prewhitening, whereas residual correlations, physiological fluctuations, and nonstationary noise may violate these assumptions. Exact risk equivalence applies to an unnormalized squared $L_2$ loss, while training used a normalized $\ell_1$--$\ell_2$ objective; the theoretical analysis should therefore be interpreted as motivation for the self-supervised design rather than an exact guarantee for the practical loss. Simulated Gaussian noise may not represent all experimental noise sources. The in vivo mouse and human experiments were feasibility studies with small sample sizes and do not establish population-level performance. Broader validation is needed across pathologies, sequences, hardware, resolutions, and undersampling factors. Computational cost, scan-specific optimization, slab-based context, and sensitivity to motion also warrant systematic evaluation.

\section{Conclusion}
We proposed a measurement-domain self-supervised framework for joint reconstruction and denoising from noisy, undersampled k-space. The method extends k-space partitioning to denoising and integrates phase regularization into an unrolled network to reduce field-induced artifacts in complex-valued GRE data. Across the evaluated high-resolution MRI datasets, SSL reduced noise while preserving fine anatomy and limiting error propagation into T$_2^*$ maps and QSM. phase-constrained self-supervision therefore provides a promising reconstruction strategy when clean high-resolution references are unavailable, while larger in vivo studies are needed to establish generalizability.

\section*{CRediT authorship contribution statement}
Juan Liu: Writing – review \& editing, Writing – original draft, Visualization, Software, Methodology, Data curation, Conceptualization. Zhuoheng Liu: Data curation. Hongbo Wu: Software. Xinyue Han: Software. Jiaen Liu: Writing – review \& editing, Supervision, Funding acquisition. Nian Wang: Writing – review \& editing, Supervision, Funding acquisition.

\section*{Declaration of competing interest}
The authors declare that they have no known competing financial interests or personal relationships that could have appeared to influence the work reported in this paper.

\section*{Disclosure of AI Use}
OpenAI ChatGPT was used to assist with English-language editing of the manuscript. The authors reviewed and take responsibility for the final content.

\section*{Acknowledgments}
This work was supported in part by the National Institutes of Health under Grant R01 NS125020, the Ralph W. and Grace M. Showalter Research Award, the UT Rising STARs Award, the Hamon Charitable Foundation, and the Texas Instruments Foundation.

\section*{Data and Code Availability}
The code supporting this study will be released in a public repository upon publication. Data will be made available from the corresponding authors on reasonable request, subject to applicable ethical and institutional restrictions.

\section*{Appendix. Supplementary materials}

\setcounter{figure}{0}
\renewcommand{\thefigure}{S\arabic{figure}}
\renewcommand{\theHfigure}{S\arabic{figure}}

\begin{figure*}
\vspace{0pt}
{\includegraphics[width=\textwidth]{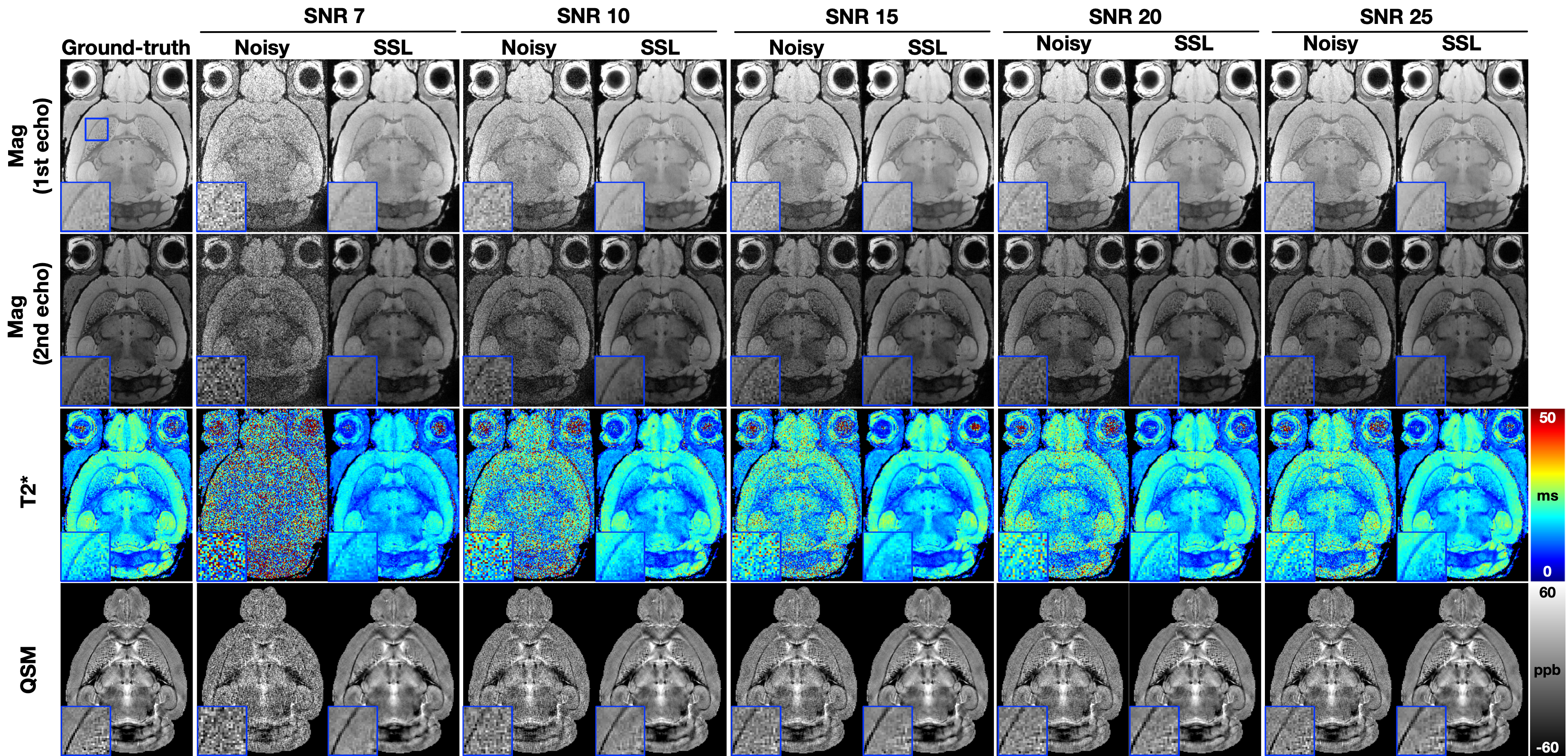}}
{\caption{Reconstruction of 72 $\mu$m isotropic dual-echo GRE ex vivo mouse brain MRI under simulated noise. Complex Gaussian noise was added to k-space to generate SNR values of 7, 10, 15, 20, and 25. The reference reconstruction had an SNR of 89 and used the original k-space data. The same noisy measurements were processed using conventional inverse Fourier reconstruction, the comparison methods, and SSL.} \label{fig:fig_ssl_72um_noiselevels}}
\end{figure*}

Figure~\ref{fig:fig_ssl_72um_noiselevels} shows representative reconstructions at five simulated noise levels. Across the displayed conditions, SSL reduced visible noise while retaining anatomical structure, although image quality decreased as the measurement SNR approached the lowest evaluated levels.

\begin{figure}
\vspace{0pt}
{\includegraphics[width=\columnwidth,height=0.8\textheight,keepaspectratio]{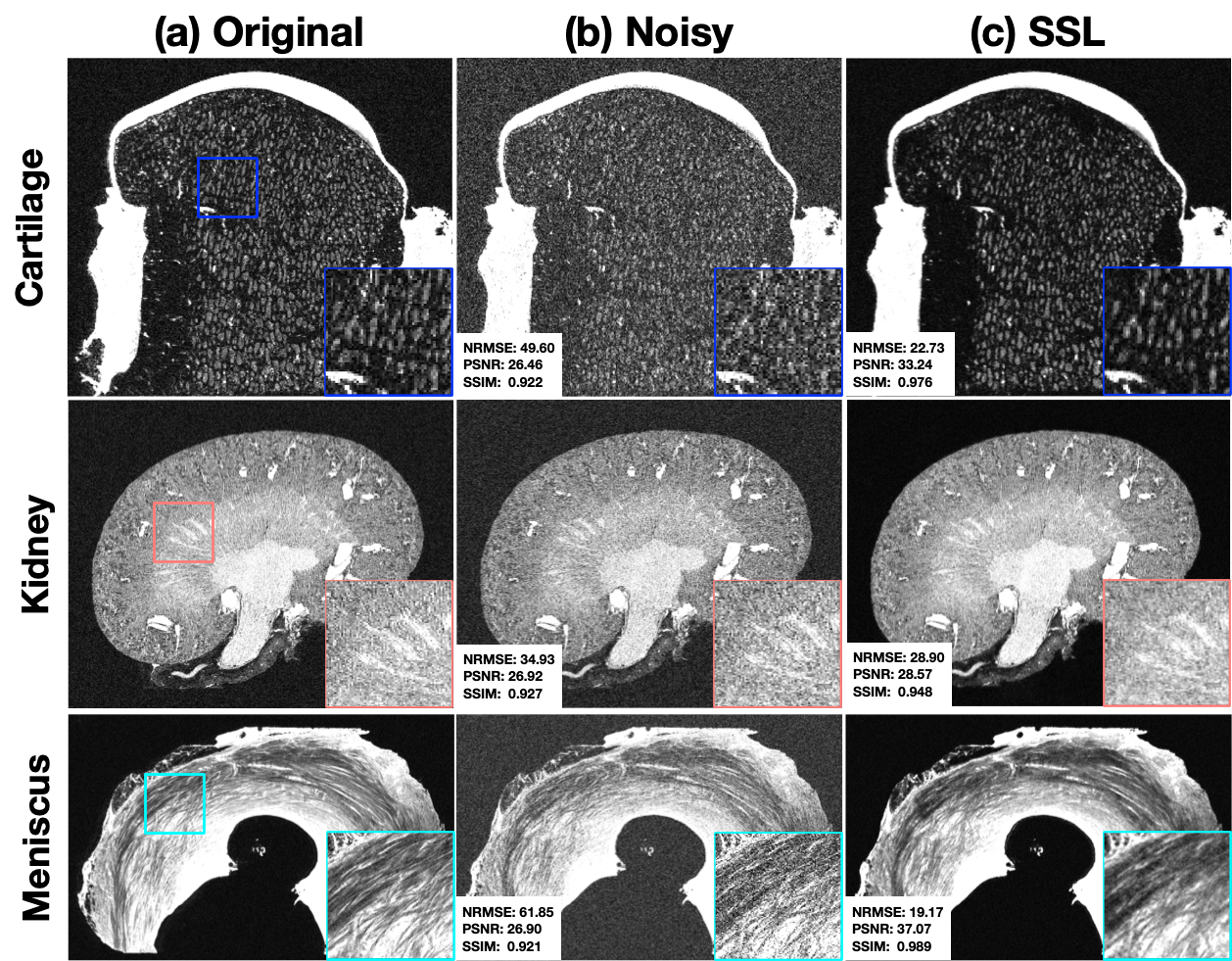}}
{\caption{Exploratory application of SSL to ex vivo mouse MRI from three organs. (a) Reconstructions from the original measurements, (b) conventional reconstructions after complex Gaussian noise was added to k-space, and (c) SSL reconstructions from the same noise-augmented measurements.}\label{fig:fig_ssl_organs}}
\end{figure}

Figure~\ref{fig:fig_ssl_organs} presents an exploratory application of SSL to ex vivo mouse MRI from three organs. After complex Gaussian noise was added to k-space, conventional reconstruction showed increased noise and reduced anatomical visibility. SSL reduced the visible noise while retaining organ-specific structures in these representative examples.

\clearpage

% References
\bibliographystyle{unsrtnat}
\bibliography{references}

@article{pruessmann1999sense,
  title={{SENSE}: sensitivity encoding for fast {MRI}},
  author={Pruessmann, Klaas P and Weiger, Markus and Scheidegger, Markus B and Boesiger, Peter},
  journal={Magnetic Resonance in Medicine: An Official Journal of the International Society for Magnetic Resonance in Medicine},
  volume={42},
  number={5},
  pages={952--962},
  year={1999},
  publisher={Wiley Online Library}
}

@article{griswold2002generalized,
  title={Generalized autocalibrating partially parallel acquisitions ({GRAPPA})},
  author={Griswold, Mark A and Jakob, Peter M and Heidemann, Robin M and Nittka, Mathias and Jellus, Vladimir and Wang, Jianmin and Kiefer, Berthold and Haase, Axel},
  journal={Magnetic Resonance in Medicine: An Official Journal of the International Society for Magnetic Resonance in Medicine},
  volume={47},
  number={6},
  pages={1202--1210},
  year={2002},
  publisher={Wiley Online Library}
}

@article{han2024high,
  title={High-resolution diffusion magnetic resonance imaging and spatial-transcriptomic in developing mouse brain},
  author={Han, Xinyue and Maharjan, Surendra and Chen, Jie and Zhao, Yi and Qi, Yi and White, Leonard E and Johnson, G Allan and Wang, Nian},
  journal={Neuroimage},
  volume={297},
  pages={120734},
  year={2024},
  publisher={Elsevier}
}

@article{serger2026motion,
  title={Motion- and Field-Robust Mesoscopic Whole-Brain {$T_2^*$}-Weighted Imaging at 7 and 11.7 T Using Servo Navigation},
  author={Serger, Matthias and Stirnberg, R{\"u}diger and Ehses, Philipp and Riedel, Malte and Ulrich, Thomas and Le Ster, Caroline and Mauconduit, Franck and Gras, Vincent and Amadon, Alexis and Vignaud, Alexandre and others},
  journal={Magnetic resonance in medicine},
  volume={95},
  number={5},
  pages={2658--2670},
  year={2026},
  publisher={Wiley Online Library}
}

@article{han2026high,
  title={High-resolution {MRI}-guided whole mouse brain neuronal cell type atlas using deep learning},
  author={Han, Xinyue and Hu, Rui and Liu, Zhuoheng and Chen, Jie and Jafry, Mubashir and Song, Haoyu and Zhao, Yi and Lin, Mingquan and White, Leonard E and Johnson, G Allan and others},
  journal={Communications Biology},
  year={2026},
  doi={10.1038/s42003-026-10608-y},
  publisher={Nature Publishing Group UK London}
}

@article{dong2025romer,
  title={Romer-{EPTI}: rotating-view motion-robust super-resolution {EPTI} for {SNR}-efficient distortion-free in-vivo mesoscale diffusion {MRI} and microstructure imaging},
  author={Dong, Zijing and Reese, Timothy G and Lee, Hong-Hsi and Huang, Susie Y and Polimeni, Jonathan R and Wald, Lawrence L and Wang, Fuyixue},
  journal={Magnetic resonance in medicine},
  volume={93},
  number={4},
  pages={1535--1555},
  year={2025},
  publisher={Wiley Online Library}
}

@article{wang2019probing,
  title={Probing demyelination and remyelination of the cuprizone mouse model using multimodality {MRI}},
  author={Wang, Nian and Zhuang, Jie and Wei, Hongjiang and Dibb, Russell and Qi, Yi and Liu, Chunlei},
  journal={Journal of Magnetic Resonance Imaging},
  volume={50},
  number={6},
  pages={1852--1865},
  year={2019},
  publisher={Wiley Online Library}
}

@article{liu2026mesoscale,
  title={Mesoscale Whole-Brain {$T_2^*$}-Weighted and Associated Quantitative {MRI} in Humans at 10.5 T},
  author={Liu, Jiaen and van Gelderen, Peter and de Zwart, Jacco A and Duyn, Jeff H and Huang, Yujia and Qu, Shuxian and Grant, Andrea and Auerbach, Edward J and Waks, Matt and Lagore, Russell L and others},
  journal={Magnetic Resonance in Medicine},
  volume={96},
  number={2},
  pages={817--825},
  year={2026},
  publisher={Wiley Online Library}
}

@article{lustig2008compressed,
  title={Compressed sensing {MRI}},
  author={Lustig, Michael and Donoho, David L and Santos, Juan M and Pauly, John M},
  journal={IEEE signal processing magazine},
  volume={25},
  number={2},
  pages={72--82},
  year={2008},
  publisher={IEEE}
}

@article{lustig2007sparse,
  title={Sparse {MRI}: The application of compressed sensing for rapid {MR} imaging},
  author={Lustig, Michael and Donoho, David and Pauly, John M},
  journal={Magnetic Resonance in Medicine: An Official Journal of the International Society for Magnetic Resonance in Medicine},
  volume={58},
  number={6},
  pages={1182--1195},
  year={2007},
  publisher={Wiley Online Library}
}

@article{yaman2020self,
  title={Self-supervised learning of physics-guided reconstruction neural networks without fully sampled reference data},
  author={Yaman, Burhaneddin and Hosseini, Seyed Amir Hossein and Moeller, Steen and Ellermann, Jutta and U{\u{g}}urbil, K{\^a}mil and Ak{\c{c}}akaya, Mehmet},
  journal={Magnetic resonance in medicine},
  volume={84},
  number={6},
  pages={3172--3191},
  year={2020},
  publisher={Wiley Online Library}
}

@inproceedings{liu2021swin,
  title={Swin transformer: Hierarchical vision transformer using shifted windows},
  author={Liu, Ze and Lin, Yutong and Cao, Yue and Hu, Han and Wei, Yixuan and Zhang, Zheng and Lin, Stephen and Guo, Baining},
  booktitle={Proceedings of the IEEE/CVF international conference on computer vision},
  pages={10012--10022},
  year={2021}
}

@inproceedings{dosovitskiy2020image,
  title={An image is worth 16x16 words: Transformers for image recognition at scale},
  author={Dosovitskiy, Alexey and Beyer, Lucas and Kolesnikov, Alexander and Weissenborn, Dirk and Zhai, Xiaohua and Unterthiner, Thomas and Dehghani, Mostafa and Minderer, Matthias and Heigold, Georg and Gelly, Sylvain and others},
  booktitle={International Conference on Learning Representations},
  year={2021}
}

@inproceedings{he2016identity,
  title={Identity mappings in deep residual networks},
  author={He, Kaiming and Zhang, Xiangyu and Ren, Shaoqing and Sun, Jian},
  booktitle={European conference on computer vision},
  pages={630--645},
  year={2016},
  organization={Springer}
}

@article{monga2021algorithm,
  title={Algorithm unrolling: Interpretable, efficient deep learning for signal and image processing},
  author={Monga, Vishal and Li, Yuelong and Eldar, Yonina C},
  journal={IEEE Signal Processing Magazine},
  volume={38},
  number={2},
  pages={18--44},
  year={2021},
  publisher={IEEE}
}

@article{feinberg2023next,
  title={Next-generation {MRI} scanner designed for ultra-high-resolution human brain imaging at 7 Tesla},
  author={Feinberg, David A and Beckett, Alexander JS and Vu, An T and Stockmann, Jason and Huber, Laurentius and Ma, Samantha and Ahn, Sinyeob and Setsompop, Kawin and Cao, Xiaozhi and Park, Suhyung and others},
  journal={Nature methods},
  volume={20},
  number={12},
  pages={2048--2057},
  year={2023},
  publisher={Nature Publishing Group US New York}
}

@article{boulant2024vivo,
  title={In vivo imaging of the human brain with the Iseult 11.7-T {MRI} scanner},
  author={Boulant, Nicolas and Mauconduit, Franck and Gras, Vincent and Amadon, Alexis and Le Ster, Caroline and Luong, Michel and Massire, Aur{\'e}lien and Pallier, Christophe and Sabatier, Laure and Bottlaender, Michel and others},
  journal={Nature Methods},
  volume={21},
  number={11},
  pages={2013--2016},
  year={2024},
  publisher={Nature Publishing Group US New York}
}

@article{rajiah2023cardiac,
  title={Cardiac {MRI}: state of the art},
  author={Rajiah, Prabhakar Shantha and Fran{\c{c}}ois, Christopher J and Leiner, Tim},
  journal={Radiology},
  volume={307},
  number={3},
  pages={e223008},
  year={2023},
  publisher={Radiological Society of North America}
}

@inproceedings{batson2019noise2self,
  title={{Noise2Self}: Blind denoising by self-supervision},
  author={Batson, Joshua and Royer, Loic},
  booktitle={International conference on machine learning},
  pages={524--533},
  year={2019},
  organization={PMLR}
}

@inproceedings{krull2019noise2void,
  title={{Noise2Void}-learning denoising from single noisy images},
  author={Krull, Alexander and Buchholz, Tim-Oliver and Jug, Florian},
  booktitle={Proceedings of the IEEE/CVF conference on computer vision and pattern recognition},
  pages={2129--2137},
  year={2019}
}

@article{liu2021self,
  title={Self-supervised learning: Generative or contrastive},
  author={Liu, Xiao and Zhang, Fanjin and Hou, Zhenyu and Mian, Li and Wang, Zhaoyu and Zhang, Jing and Tang, Jie},
  journal={IEEE transactions on knowledge and data engineering},
  volume={35},
  number={1},
  pages={857--876},
  year={2021},
  publisher={IEEE}
}

@article{plenge2012super,
  title={Super-resolution methods in {MRI}: can they improve the trade-off between resolution, signal-to-noise ratio, and acquisition time?},
  author={Plenge, Esben and Poot, Dirk HJ and Bernsen, Monique and Kotek, Gyula and Houston, Gavin and Wielopolski, Piotr and Van Der Weerd, Louise and Niessen, Wiro J and Meijering, Erik},
  journal={Magnetic resonance in medicine},
  volume={68},
  number={6},
  pages={1983--1993},
  year={2012},
  publisher={Wiley Online Library}
}

@inproceedings{sun2016deep,
  title={Deep {ADMM-Net} for compressive sensing {MRI}},
  author={Yang, Yan and Sun, Jian and Li, Huibin and Xu, Zongben},
  booktitle={Advances in Neural Information Processing Systems},
  volume={29},
  pages={10--18},
  year={2016}
}

@article{frisoni2010clinical,
  title={The clinical use of structural {MRI} in Alzheimer disease},
  author={Frisoni, Giovanni B and Fox, Nick C and Jack Jr, Clifford R and Scheltens, Philip and Thompson, Paul M},
  journal={Nature reviews neurology},
  volume={6},
  number={2},
  pages={67--77},
  year={2010},
  publisher={Nature Publishing Group UK London}
}

@article{mahlknecht2010significance,
  title={Significance of {MRI} in diagnosis and differential diagnosis of Parkinson’s disease},
  author={Mahlknecht, Philipp and Hotter, Anna and Hussl, Anna and Esterhammer, Regina and Schocke, Michael and Seppi, Klaus},
  journal={Neurodegenerative Diseases},
  volume={7},
  number={5},
  pages={300--318},
  year={2010},
  publisher={S. Karger AG Basel, Switzerland}
}

@article{sneag2023mri,
  title={{MRI} advancements in musculoskeletal clinical and research practice},
  author={Sneag, Darryl B and Abel, Frederik and Potter, Hollis G and Fritz, Jan and Koff, Matthew F and Chung, Christine B and Pedoia, Valentina and Tan, Ek T},
  journal={Radiology},
  volume={308},
  number={2},
  pages={e230531},
  year={2023},
  publisher={Radiological Society of North America}
}

@article{vosshenrich2024modern,
  title={Modern acceleration in musculoskeletal {MRI}: applications, implications, and challenges},
  author={Vosshenrich, Jan and Koerzdoerfer, Gregor and Fritz, Jan},
  journal={Skeletal Radiology},
  volume={53},
  number={9},
  pages={1799--1813},
  year={2024},
  publisher={Springer}
}

@article{morales2024present,
  title={Present and future innovations in {AI} and cardiac {MRI}},
  author={Morales, Manuel A and Manning, Warren J and Nezafat, Reza},
  journal={Radiology},
  volume={310},
  number={1},
  pages={e231269},
  year={2024},
  publisher={Radiological Society of North America}
}

@article{ng2023mri,
  title={{MRI}-{LINAC}: A transformative technology in radiation oncology},
  author={Ng, John and Gregucci, Fabiana and Pennell, Ryan T and Nagar, Himanshu and Golden, Encouse B and Knisely, Jonathan PS and Sanfilippo, Nicholas J and Formenti, Silvia C},
  journal={Frontiers in Oncology},
  volume={13},
  pages={1117874},
  year={2023},
  publisher={Frontiers Media SA}
}

@article{lagendijk2008mri,
  title={{MRI}/linac integration},
  author={Lagendijk, Jan JW and Raaymakers, Bas W and Raaijmakers, Alexander JE and Overweg, Johan and Brown, Kevin J and Kerkhof, Ellen M and van der Put, Richard W and H{\aa}rdemark, Bj{\"o}rn and van Vulpen, Marco and van der Heide, Uulke A},
  journal={Radiotherapy and Oncology},
  volume={86},
  number={1},
  pages={25--29},
  year={2008},
  publisher={Elsevier}
}

@article{bakshi2008mri,
  title={{MRI} in multiple sclerosis: current status and future prospects},
  author={Bakshi, Rohit and Thompson, Alan J and Rocca, Maria A and Pelletier, Daniel and Dousset, Vincent and Barkhof, Frederik and Inglese, Matilde and Guttmann, Charles RG and Horsfield, Mark A and Filippi, Massimo},
  journal={The Lancet Neurology},
  volume={7},
  number={7},
  pages={615--625},
  year={2008},
  publisher={Elsevier}
}

@inproceedings{he2022masked,
  title     = {Masked Autoencoders Are Scalable Vision Learners},
  author    = {He, Kaiming and Chen, Xinlei and Xie, Saining and Li, Yanghao and Doll{\'a}r, Piotr and Girshick, Ross},
  booktitle = {Proceedings of the IEEE/CVF Conference on Computer Vision and Pattern Recognition (CVPR)},
  pages     = {15979--15988},
  year      = {2022}
}

@article{buades2011non,
  title={Non-local means denoising},
  author={Buades, Antoni and Coll, Bartomeu and Morel, Jean-Michel},
  journal={Image processing on line},
  volume={1},
  pages={208--212},
  year={2011}
}

@inproceedings{huang2021neighbor2neighbor,
  title={Neighbor2neighbor: Self-supervised denoising from single noisy images},
  author={Huang, Tao and Li, Songjiang and Jia, Xu and Lu, Huchuan and Liu, Jianzhuang},
  booktitle={Proceedings of the IEEE/CVF conference on computer vision and pattern recognition},
  pages={14781--14790},
  year={2021}
}

@inproceedings{uecker2015berkeley,
  title={Berkeley advanced reconstruction toolbox},
  author={Uecker, Martin and Ong, Frank and Tamir, Jonathan I and Bahri, Dara and Virtue, Patrick and Cheng, Joseph Y and Zhang, Tao and Lustig, Michael},
  booktitle={Proc. Intl. Soc. Mag. Reson. Med},
  volume={23},
  number={2486},
  pages={9},
  year={2015}
}

@article{polak2020nonlinear,
  title={Nonlinear dipole inversion ({NDI}) enables robust quantitative susceptibility mapping ({QSM})},
  author={Polak, Daniel and Chatnuntawech, Itthi and Yoon, Jaeyeon and Iyer, Siddharth Srinivasan and Milovic, Carlos and Lee, Jongho and Bachert, Peter and Adalsteinsson, Elfar and Setsompop, Kawin and Bilgic, Berkin},
  journal={NMR in Biomedicine},
  volume={33},
  number={12},
  pages={e4271},
  year={2020},
  publisher={Wiley Online Library}
}

@article{uecker2014espirit,
  title={{ESPIRiT}—an eigenvalue approach to autocalibrating parallel {MRI}: where {SENSE} meets {GRAPPA}},
  author={Uecker, Martin and Lai, Peng and Murphy, Mark J and Virtue, Patrick and Elad, Michael and Pauly, John M and Vasanawala, Shreyas S and Lustig, Michael},
  journal={Magnetic resonance in medicine},
  volume={71},
  number={3},
  pages={990--1001},
  year={2014},
  publisher={Wiley Online Library}
}

@article{bernstein2001effect,
  title={Effect of windowing and zero-filled reconstruction of {MRI} data on spatial resolution and acquisition strategy},
  author={Bernstein, Matt A and Fain, Sean B and Riederer, Stephen J},
  journal={Journal of magnetic resonance imaging: an official journal of the international society for magnetic resonance in medicine},
  volume={14},
  number={3},
  pages={270--280},
  year={2001},
  publisher={Wiley Online Library}
}

@article{rudin1992nonlinear,
  title={Nonlinear total variation based noise removal algorithms},
  author={Rudin, Leonid I and Osher, Stanley and Fatemi, Emad},
  journal={Physica D: nonlinear phenomena},
  volume={60},
  number={1-4},
  pages={259--268},
  year={1992},
  publisher={Elsevier}
}

@article{chung2022score,
  title={Score-based diffusion models for accelerated {MRI}},
  author={Chung, Hyungjin and Ye, Jong Chul},
  journal={Medical image analysis},
  volume={80},
  pages={102479},
  year={2022},
  publisher={Elsevier}
}

@inproceedings{song2021solving,
  title={Solving inverse problems in medical imaging with score-based generative models},
  author={Song, Yang and Shen, Liyue and Xing, Lei and Ermon, Stefano},
  booktitle={International Conference on Learning Representations},
  year={2022}
}

@article{van2023effect,
  title={Effect of motion, cortical orientation and spatial resolution on quantitative imaging of cortical {$R_2^*$} and magnetic susceptibility at 0.3 mm in-plane resolution at {7T}},
  author={van Gelderen, Peter and Li, Xu and de Zwart, Jacco A and Beck, Erin S and Okar, Serhat V and Huang, Yujia and Lai, KuoWei and Sulam, Jeremias and van Zijl, Peter CM and Reich, Daniel S and others},
  journal={Neuroimage},
  volume={270},
  pages={119992},
  year={2023},
  publisher={Elsevier}
}

@article{ozbay2017comprehensive,
  title={A comprehensive numerical analysis of background phase correction with {V-SHARP}},
  author={{\"O}zbay, Pinar Senay and Deistung, Andreas and Feng, Xiang and Nanz, Daniel and Reichenbach, J{\"u}rgen Rainer and Schweser, Ferdinand},
  journal={NMR in Biomedicine},
  volume={30},
  number={4},
  pages={e3550},
  year={2017},
  publisher={Wiley Online Library}
}

@article{schofield2003fast,
  title={Fast phase unwrapping algorithm for interferometric applications},
  author={Schofield, Marvin A and Zhu, Yimei},
  journal={Optics letters},
  volume={28},
  number={14},
  pages={1194--1196},
  year={2003},
  publisher={Optical Society of America}
}

@inproceedings{frangi1998multiscale,
  title={Multiscale vessel enhancement filtering},
  author={Frangi, Alejandro F and Niessen, Wiro J and Vincken, Koen L and Viergever, Max A},
  booktitle={International conference on medical image computing and computer-assisted intervention},
  pages={130--137},
  year={1998},
  organization={Springer}
}

@article{quah2026complex,
  title={Complex deep neural networks for denoising ultra-fast submillimeter {$T_2^*$}-weighted imaging and quantitative susceptibility mapping},
  author={Quah, Bryan and Nair, Sreekanth Madhusoodhanan and Gao, Chang and Luu, Michael and Renner, Brian and Jin, Jin and Han, Fei and Luskin, Elaina and Silemek, Arzu C Has and Binesh, Nader and others},
  journal={European Journal of Radiology Artificial Intelligence},
  volume={6},
  pages={100100},
  year={2026},
  publisher={Elsevier}
}

@article{zhang2022blip,
  title={Blip up-down acquisition for spin-and gradient-echo imaging ({BUDA-SAGE}) with self-supervised denoising enables efficient {$T_2$}, {$T_2^*$}, para-and dia-magnetic susceptibility mapping},
  author={Zhang, Zijing and Cho, Jaejin and Wang, Long and Liao, Congyu and Shin, Hyeong-Geol and Cao, Xiaozhi and Lee, Jongho and Xu, Jinmin and Zhang, Tao and Ye, Huihui and others},
  journal={Magnetic Resonance in Medicine},
  volume={88},
  number={2},
  pages={633--650},
  year={2022},
  publisher={Wiley Online Library}
}

@article{veraart2016denoising,
  title={Denoising of diffusion {MRI} using random matrix theory},
  author={Veraart, Jelle and Novikov, Dmitry S and Christiaens, Daan and Ades-Aron, Benjamin and Sijbers, Jan and Fieremans, Els},
  journal={Neuroimage},
  volume={142},
  pages={394--406},
  year={2016},
  publisher={Elsevier}
}

@article{fadnavis2020patch2self,
  title={{Patch2Self}: Denoising diffusion {MRI} with self-supervised learning},
  author={Fadnavis, Shreyas and Batson, Joshua and Garyfallidis, Eleftherios},
  journal={Advances in neural information processing systems},
  volume={33},
  pages={16293--16303},
  year={2020}
}

@article{zormpas2021accelerating,
  title={Accelerating whole-body diffusion-weighted {MRI} with deep learning--based denoising image filters},
  author={Zormpas-Petridis, Konstantinos and Tunariu, Nina and Curcean, Andra and Messiou, Christina and Curcean, Sebastian and Collins, David J and Hughes, Julie C and Jamin, Yann and Koh, Dow-Mu and Blackledge, Matthew D},
  journal={Radiology: Artificial Intelligence},
  volume={3},
  number={5},
  pages={e200279},
  year={2021},
  publisher={Radiological Society of North America}
}

@article{manjon2008mri,
  title={{MRI} denoising using non-local means},
  author={Manj{\'o}n, Jos{\'e} V and Carbonell-Caballero, Jos{\'e} and Lull, Juan J and Garc{\'\i}a-Mart{\'\i}, Graci{\'a}n and Mart{\'\i}-Bonmat{\'\i}, Lu{\'\i}s and Robles, Montserrat},
  journal={Medical image analysis},
  volume={12},
  number={4},
  pages={514--523},
  year={2008},
  publisher={Elsevier}
}

@article{kang2024self,
  title={Self-supervised learning for denoising of multidimensional {MRI} data},
  author={Kang, Beomgu and Lee, Wonil and Seo, Hyunseok and Heo, Hye-Young and Park, HyunWook},
  journal={Magnetic resonance in medicine},
  volume={92},
  number={5},
  pages={1980--1994},
  year={2024},
  publisher={Wiley Online Library}
}

@inproceedings{fadnavis2024patch2self2,
  title={{Patch2Self2}: Self-supervised denoising on coresets via matrix sketching},
  author={Fadnavis, Shreyas and Chowdhury, Agniva and Batson, Joshua and Drineas, Petros and Garyfallidis, Eleftherios},
  booktitle={Proceedings of the IEEE/CVF Conference on Computer Vision and Pattern Recognition},
  pages={27641--27651},
  year={2024}
}

@article{koonjoo2021boosting,
  title={Boosting the signal-to-noise of low-field {MRI} with deep learning image reconstruction},
  author={Koonjoo, Neha and Zhu, Bo and Bagnall, G Cody and Bhutto, D and Rosen, MS},
  journal={Scientific reports},
  volume={11},
  number={1},
  pages={8248},
  year={2021},
  publisher={Nature Publishing Group UK London}
}

@article{desai2023noise2recon,
  title={{Noise2Recon}: Enabling {SNR}-robust {MRI} reconstruction with semi-supervised and self-supervised learning},
  author={Desai, Arjun D and Ozturkler, Batu M and Sandino, Christopher M and Boutin, Robert and Willis, Marc and Vasanawala, Shreyas and Hargreaves, Brian A and R{\'e}, Christopher and Pauly, John M and Chaudhari, Akshay S},
  journal={Magnetic Resonance in Medicine},
  volume={90},
  number={5},
  pages={2052--2070},
  year={2023},
  publisher={Wiley Online Library}
}

@article{millard2023theoretical,
  title={A theoretical framework for self-supervised {MR} image reconstruction using sub-sampling via variable density {Noisier2Noise}},
  author={Millard, Charles and Chiew, Mark},
  journal={IEEE transactions on computational imaging},
  volume={9},
  pages={707--720},
  year={2023},
  publisher={IEEE}
}

@article{chen2026self,
  title={Self-supervised joint reconstruction and denoising of {$T_2$}-Weighted {PROPELLER} {MRI} of the lung at 0.55 T},
  author={Chen, Jingjia and Pei, Haoyang and Maier, Christoph and Bruno, Mary and Wen, Qiuting and Shin, Seon-Hi and Moore, William and Chandarana, Hersh and Feng, Li},
  journal={Magnetic resonance in medicine},
  volume={95},
  number={5},
  pages={2930--2946},
  year={2026},
  publisher={Wiley Online Library}
}

@inproceedings{moran2020noisier2noise,
  title={Noisier2noise: Learning to denoise from unpaired noisy data},
  author={Moran, Nick and Schmidt, Dan and Zhong, Yu and Coady, Patrick},
  booktitle={Proceedings of the IEEE/CVF conference on computer vision and pattern recognition},
  pages={12064--12072},
  year={2020}
}
\end{document}